\documentclass{llncs}

\usepackage{amsmath}
\usepackage{graphicx}
\usepackage{xcolor}
\usepackage{xspace}
\usepackage{proof}
\usepackage{epsfig}
\usepackage{graphicx}
\usepackage{tabularx}

\usepackage{txfonts} 

\usepackage{color} 
\definecolor{blue}{rgb}{0.2,0.2,0.4}
\definecolor{brown}{rgb}{0.7,0.4,0.3}
\definecolor{red}{rgb}{1,0.2,0}
\definecolor{yellow}{rgb}{1,1,0}

\newcommand{\LinearLogic}{Linear Logic\xspace}
\newcommand{\Sequences}{Sequences\xspace}
\newcommand{\Sequence}{Sequence\xspace}
\newcommand{\sequence}{sequence\xspace}
\newcommand{\BigSteps}{Big Steps\xspace}

\newcommand{\lcLamCalc}{$\lambda$-calculus\xspace}
\newcommand{\lcLamTerm}{$\lambda$-term\xspace}
\newcommand{\lcLamTerms}{$\lambda$-terms\xspace}

\newcommand{\polytime}{polynomial time\xspace}

\newcommand{\Size}[1]{|#1|} 
\newcommand{\Seq}[2]{ #1 \ldots #2  } 
\newcommand{\Set}[1]{ \{ #1 \}} 
\newcommand{\SetSq}[1]{ [ #1 ]} 
\newcommand{\BNFDef}{::=}
\newcommand{\Def}[2]{#1 \equiv #2}

\newcommand{\TFA}{\textsf{TFA}\xspace}
\newcommand{\LAL}{\textsf{LAL}\xspace}
\newcommand{\DLAL}{\textsf{DLAL}\xspace}
\newcommand{\SLL}{\textsf{SLL}\xspace}

\newcommand{\lImpl}{\!\multimap\!}
\newcommand{\lTime}{\!\otimes\!}
\newcommand{\lPar}{\S}

\newcommand{\lBang}{\,!}
\newcommand{\hasType}{:}

\newcommand{\lcFV}[1]{\operatorname{fv}(#1)} 
\newcommand{\lcSub}[2]{\Set{^{#1}\!/_{#2}}} 
\newcommand{\lcSubT}[4]{\Set{^{#1}\!/_{#2}\, ^{#3}\!/_{#4}}} 
\newcommand{\lcSubM}[4]{\Set{^{#1}\!/_{#2}, \ldots ,^{#3}\!/_{#4}}} 
\newcommand{\lcVar}{\mathcal{V}} %
\newcommand{\lcSet}{\Lambda} %
\newcommand{\lcSetVal}{\Lambda^{\!\textrm{v}}} %
\newcommand{\lcSetNF}{\Lambda^{\!\textrm{v}}} %
\newcommand{\lcF}[2]{\lambda #1.#2} 
\newcommand{\lcA}[2]{(#1)\, #2} 
\newcommand{\lcPC}[2]{\langle #1, #2 \rangle} 
\newcommand{\lcPD}[3]{\lcF{\langle #1, #2 \rangle}{#3}} 
\newcommand{\lcTC}[1]{\langle #1  \rangle} 
\newcommand{\lcTD}[2]{\lcF{\langle #1  \rangle}{#2}} 
\newcommand{\lcId}{\lcF{x}{x}} %

\newcommand{\lcNFJud}[2]{ #1 \Downarrow #2} %
\newcommand{\lcNFv}{\textrm{\small v}} %
\newcommand{\lcNFf}{\textrm{\small f}} %
\newcommand{\lcNFp}{\textrm{\small p}} %
\newcommand{\lcNFap}{\textrm{\small @p}} %
\newcommand{\lcNFava}{\textrm{\small @v}} %
\newcommand{\lcNFavl}{\textrm{\small @l}} %
\newcommand{\taFV}[1]{\operatorname{fv}(#1)} 
\newcommand{\taSub}[2]{\SetSq{^{#1}\!/_{#2}}} 
\newcommand{\taVar}{\mathcal{G}} 
\newcommand{\taFor}{\mathcal{F}} 
\newcommand{\taTyp}{\mathcal{T}} 
\newcommand{\taJudg}[4]{ #1 \mid #2 \vdash #3 \!\hasType\! #4}
\newcommand{\taCntxE}{\Delta} %
\newcommand{\taCntxL}{\Gamma} %
\newcommand{\taCntxEmpty}{\emptyset} %
\newcommand{\taCntxB}[2]{ #1\!\hasType\!#2} %
\newcommand{\taAx}{\textrm{\small a}} 
\newcommand{\taW}{\textrm{\small w}} 
\newcommand{\taC}{\textrm{\small c}} 
\newcommand{\taLII}{\lImpl \textrm{\small I}} 
\newcommand{\taLIE}{\lImpl \textrm{\small E}} 
\newcommand{\taLIIp}{\lImpl \textrm{\small I}_{\ \lTime}} 
\newcommand{\taTup}{\lTime\ \textrm{\small I}} 
\newcommand{\taEII}{\Rightarrow\!\textrm{\small I}} 
\newcommand{\taEIE}{\Rightarrow\!\!\textrm{\small E}} 
\newcommand{\taQI }{\forall \textrm{\small I}} 
\newcommand{\taQE }{\forall \textrm{\small E}} 
\newcommand{\taPI }{\lPar \textrm{\small I}} 
\newcommand{\taPE }{\lPar \textrm{\small E}} 
\newcommand{\taPL}{\lPar\textrm{\small L}} 

\newcommand{\lcBL}[1]{\lBang[#1]} 
\newcommand{\lcPL}[2]{\lPar^{#1}[#2]} 

\newcommand{\pTupC}[1]{\langle #1 \rangle} %
\newcommand{\tCast}[1]{\mathtt{tCast}^{#1}} %

\newcommand{\bBit}[1]{{\mathtt b_{#1}}} %
\newcommand{\bBitP}[1]{{\mathtt b'_{#1}}} %
\newcommand{\bBitPP}[1]{{\mathtt b''_{#1}}} %
\newcommand{\bBitr}[1]{{\mathtt r_{#1}}} 
\newcommand{\bBitrP}[1]{{\mathtt r'_{#1}}} 
\newcommand{\bBita}[1]{{\mathtt a_{#1}}} 
\newcommand{\bBitc}[1]{{\mathtt c_{#1}}} %
\newcommand{\bBits}[1]{{\mathtt s_{#1}}} 
\newcommand{\bBitg}[1]{{\mathtt g_{#1}}} 
\newcommand{\bBitm}[1]{{\mathtt m_{#1}}} 
\newcommand{\bBitp}[1]{{\mathtt p_{#1}}} 
\newcommand{\bBitM}[1]{{\mathtt M_{#1}}} 
\newcommand{\bBitd}[1]{{\mathtt d_{#1}}} 
\newcommand{\bTT}{\mathtt{1}} %
\newcommand{\bFF}{\mathtt{0}} %
\newcommand{\bBot}{\mathtt{\perp}} %
\newcommand{\bXor}{\mathtt{Xor}} %
\newcommand{\bAnd}{\mathtt{And}} %
\newcommand{\blsb}{l.s.b.\xspace} %
\newcommand{\bmsb}{m.s.b.\xspace} %
\newcommand{\bDup}[1]{\mathtt{b}\!\nabla_{\!#1}} %
\newcommand{\bCast}[1]{\mathtt{bCast}^{#1}} %

\newcommand{\sExt}[1]{ [ #1 ] } %
\newcommand{\sNil}{\sExt{\varepsilon}} %
\newcommand{\sSplit}{\mathtt{sSpl}} %

\newcommand{\Cast}[1]{\mathtt{Cast}^{#1}} %
\newcommand{\wExt}[1]{ \{ #1 \} } %
\newcommand{\wCast}[1]{\mathtt{wCast}^{#1}} %
\newcommand{\wShiftRBot}{\mathtt{wShiftR}\bBot} %
\newcommand{\wShiftRBotStep}[1]{\mathtt{wShiftR}\bBot\mathtt{Step}[#1]} %
\newcommand{\wDup}[2]{\mathtt{w}\nabla_{#1}^{#2}} %
\newcommand{\wDupStep}{\mathtt{w}\nabla\mathtt{Step}} %
\newcommand{\wDupBase}{\mathtt{w}\nabla\mathtt{Base}} %
\newcommand{\wNil}{\wExt{\varepsilon}} %
\newcommand{\wSuc}[1]{\mathtt{wSuc}#1} %
\newcommand{\wSucZ}{\mathtt{wSuc}\bFF} %
\newcommand{\wSucO}{\mathtt{wSuc}\bTT} %
\newcommand{\wSucBot}{\mathtt{wSuc}\bBot} %
\newcommand{\wRev}{\mathtt{wRev}} %
\newcommand{\wRevStep}[1]{\mathtt{wRevStep}[#1]} %
\newcommand{\wDropB}{\mathtt{wDrop\bBot}} %
\newcommand{\wTos}{\mathtt{w2s}} %
\newcommand{\wTosFromFold}{\mathtt{w2sFromFold}} %
\newcommand{\wTou}{\mathtt{w2u}} %
\newcommand{\wMap}[1]{\mathtt{Map}[#1]} %
\newcommand{\wMapFromFold}[1]{\mathtt{MapFromFold}[#1]} %
\newcommand{\wFold}[2]{\mathtt{Fold}[#1,#2]} %
\newcommand{\wMapPar}{\mathtt{F}} %
\newcommand{\wMapParS}{\mathtt{S}} %
\newcommand{\wXorBW}{\mathtt{wXor}} %
\newcommand{\wLenghtDiff}{\mathtt{wLenghtDiff}} %
\newcommand{\wMT}[1]{\mathtt{MapThread}[#1]} %
\newcommand{\wMTStep}[1]{\mathtt{MTStep}[#1]} %
\newcommand{\wMTBase}[1]{\mathtt{MTBase}[#1]} %
\newcommand{\wProj}{\mathtt{wProj}} %
\newcommand{\wProjb}{\mathtt{wProj2}} %
\newcommand{\wProjFromMap}{\mathtt{wProjFromMap}} %
\newcommand{\wMS}[1]{\mathtt{MapState}[#1]} 
\newcommand{\wMSStep}[1]{\mathtt{MSStep}[#1]} %
\newcommand{\wMSBase}[1]{\mathtt{MSBase}[#1]} %

\newcommand{\wInitBot}{\mathtt{wInit}\bBot} %
\newcommand{\wMult}{\mathtt{wMult}} %
\newcommand{\wFMult}{\mathtt{wFMult}} %
\newcommand{\wBMult}[1]{\mathtt{wBMult}[#1]} %
\newcommand{\wMultStep}{\mathtt{wMultStep}} 
\newcommand{\wMultBase}{\mathtt{wMultBase}} 

\newcommand{\uExt}[1]{ \overline{#1} } %
\newcommand{\uNil}{\mathtt{u}\varepsilon} %
\newcommand{\uSuc}{\mathtt{uSuc}} %
\newcommand{\uDup}[2]{\mathtt{u}\nabla_{#1}^{#2}} %
\newcommand{\uDupStep}{\mathtt{uDupStep}} %
\newcommand{\uDupBase}{\mathtt{uDupBase}} %
\newcommand{\uDiff}{\mathtt{uDiff}} %
\newcommand{\uPred}{\mathtt{uPred}} %
\newcommand{\uPredStep}[1]{\mathtt{uPredStep}[#1]} %
\newcommand{\uPredBase}[1]{\mathtt{uPredBase}[#1]} %
\newcommand{\uCast}[1]{\mathtt{uCast}^{#1}} %

\newcommand{\pType}[4]{(#1 #2 #3)[#4]} %
\newcommand{\pTypeC}[3]{(#1 #2 #3)} 
\newcommand{\bType}[2]{\mathbb{B}_{#1}[#2]} %
\newcommand{\bTypeC}[1]{\mathbb{B}_{#1}} 
\newcommand{\sType}[1]{\mathbb{S}[#1]} %
\newcommand{\sTypeC}{\mathbb{S}} 
\newcommand{\lTypeG}[2]{\mathbb{L}(#1)[#2]}
\newcommand{\lType}[1]{\mathbb{L}(#1)}
\newcommand{\wTypeC}[1]{\mathbb{L}_{#1}} 
\newcommand{\wTypeG}[2]{\mathbb{L}(#1)[#2]}
\newcommand{\uType}[1]{\mathbb{U}[#1]} %
\newcommand{\uTypeC}{\mathbb{U}} 

\newcommand{\F}{\mathbb{F}} %
\newcommand{\bfType}{\mathbb{F}_{2^n}} %
\newcommand{\bfAdd}{\mathtt{Add}} %
\newcommand{\bfSqr}{\mathtt{Sqr}} %
\newcommand{\bfMult}{\mathtt{Mult}} %
\newcommand{\wSqr}{\mathtt{wSqr}} %
\newcommand{\wSqrStep}{\mathtt{wSqrStep}} %
\newcommand{\wMod}[2]{\mathtt{wMod[#1,#2]}} %
\newcommand{\wModParN}{\boldsymbol{\mathnormal{n}}} %
\newcommand{\wModParF}{\boldsymbol{\mathnormal{p}}} %
\newcommand{\wModEnd}{\mathtt{wModEnd}} %
\newcommand{\wModBase}{\mathtt{wModBase}} %
\newcommand{\wModFun}{\mathtt{wModFun}} %
\newcommand{\bfAPI}{\mathtt{API}} %

\newcommand{\ppMap}[1]{\mathtt{Map}[#1]} %
\newcommand{\ppMapFromFold}[1]{\mathtt{MapFromFold}[#1]}
\begin{document}

\title{Typing a Core Binary Field Arithmetic in a Light Logic}

\author{
Emanuele Cesena\inst{1}\thanks{Partially supported by the European Project TClouds (\url{http://www.tclouds-project.eu}).}
\and
Marco Pedicini\inst{2}
\and
Luca Roversi\inst{3}
}

\institute{
Politecnico di Torino, 
Dip. di Automatica e Informatica, Torino, Italy\\
\email{emanuele.cesena@polito.it}
\and
Istituto per le Applicazioni del Calcolo ``Mauro Picone'',
CNR, Roma, Italy and\\ 
LIPN -- 
UMR CNRS 7030,
Institut Galil\'ee - Universit\'e Paris-Nord \\
\email{marco@iac.cnr.it}
\and
Universit{\`a} degli Studi di Torino, 
Dip. di Informatica, Torino, Italy\\
\email{roversi@di.unito.it}
}

\maketitle

\begin{abstract}
We design a library for binary field arithmetic and we supply a core $\bfAPI$
which is completely developed in \DLAL, extended with a fix point formula. 
Since \DLAL is a restriction of linear logic where only functional programs 
with polynomial evaluation cost can be typed, we obtain the core of a 
functional programming setting for binary field arithmetic with 
\emph{built-in} polynomial complexity.
\end{abstract}

\section{Introduction}
Embedded systems (smart cards, mobile phones, sensors) are very heavy on
resources. 
Low memory and computational power force programmers to
choose specific algorithms and fine tune them in order to carefully manage 
the space and time complexity.
There is an applicative domain where these constraints on resources cause serious
difficulties: the implementation of cryptographic primitives, that is the
foundation for strong security mechanisms and protocols.
\par

We have started reasoning about a controlled programming setting, that should
enable the certification of resource usage (memory and computation time), in a
functional programming language. We are aware
of different approaches to solve analogous problems,
for instance the Computer Aided Cryptography Engineering (CACE) 
European project\footnote{\url{http://www.cace-project.eu}} whose mission is
``to enable verifiable secure cryptographic software engineering to non-experts 
by developing a toolbox which automatically produces high-performance solutions 
from natural specifications''. 
\par
What if difficulties on time/space complexity were \emph{automagically} overcome by imposing an
appropriate type discipline to the programming language?
\par
This paper describes preliminary results on how to devise a programming 
language that grants a natural programming style in the implementation 
of specific number theoretic algorithms, in combination with a type 
discipline which ensures complexity bounds. 
More precisely, we investigate how to achieve implementation of number theoretic
algorithms with certified running time bounds by exploiting logical tools under the
prescriptions of Implicit Computational Complexity (ICC) \cite{GirardLLL}.
We recall that ICC mainly aims at searching strong mathematical roots for
computational complexity theory.
The logical approach to ICC extracts functional language primitives from logical
systems under the Curry-Howard analogy.
The logical system for ICC we focus on is \DLAL~\cite{BaillotT09}.
It derives from linear logic. Its formulas can be types of \lcLamTerms. A \lcLamTerm\
$M$ typable in \DLAL\ reduces to its normal form in a time which is a polynomial in
the dimension of $M$.            
\par
We propose to put this theory into practice by developing and implementing a
\emph{core library} of combinators, namely \lcLamTerms, typeable in \DLAL.
The library currently implements a subset of functionalities which are needed
for binary field arithmetic (cf., e.g.,~\cite[Section 11.2]{avanzi05handbook}).
The practical relevance of completing such a library is to import
functional programming technology with a known predetermined complexity into
the area of applied cryptography.
\subsubsection{Contributions.}
Defining a core library that correctly implements finite field arithmetic
is a result in itself. The reason is that when programming non
obvious combinators typeable in \DLAL, the main obstacle lies in the application
of the standard \textit{divide-et-impera} paradigm: first \emph{split} the problem into
successively simpler ones until the solution becomes trivial, then \emph{compose} the
results. Composition is the harmful activity as soon as we face complexity
issues. For example, using the output of a sub-problem, which results from an
iteration, as the input of another iteration may yield a computational complexity
blowup. This is why, in \DLAL, naively manipulating lists by means of
iterations, can rapidly ``degrade'' to situations where compositions which would
be natural in standard \lcLamCalc simply get forbidden. It is for this reason that 
\lcLamTerms in \DLAL which implement the low level library with finite field operations are
not the natural ones that we could write using \lcLamTerms\ typeable in the System F
\cite{Girard-Lafont-Taylor:1989-PAT}.
\par
To overcome the need of programming with non natural \lcLamTerms, 
we follow \cite{DBLP:journals/jfp/Hutton99}, which promotes 
standard programming patterns to assure readability and soundness of functional
programs. We build an experimental $\bfAPI$ on top of the core library, which
exports standard programming patterns. The goal of supplying 
an $\bfAPI$ is to help non experts writing \lcLamTerms which are not directly
typeable in \DLAL, but which, roughly speaking, can be checked to compile into
\lcLamTerms with a type in \DLAL.
\par
\newcommand{\remove}[1]{}
\remove{To illustrate the intended use of the available patterns that the $\bfAPI$ exports,
let $\lType{A},
\lType{B}$ be the types of lists with elements of type $A, B$, and $A\lImpl B$
be the type of a linear function from $A$ to $B$.
\par
As an example, we detail how using the programming pattern $\wFold{F}{S}$
which we declare of type $\lType{A}\lImpl B$, written
$\wFold{F}{S}\hasType\lType{A}\lImpl B$, whenever $F\hasType A\lImpl
B\lImpl B$, and $S\hasType B$, and $B$ is a \emph{linear} type that admits
coerce. Being $B$ ``linear''
means $B\not\equiv \lPar C$, for any $C$. We recall that $\lPar$ is the \DLAL
type operator that marks every output obtained as conclusion of an iteration.
``Admits coerce means that the type $B\lImpl \lPar B$ is a theorem of \DLAL.
As usual, the parameter $S$ is the initial state, and $F$ the folding step that
incrementally produces the final state both from the current one and from an
element in the list.
\par
We allow a programmer to exploit $\wFold{\cdot}{\cdot}$ by \emph{proposing}
extensions of the $\bfAPI$. Every proposed extension must be a
definition like $\texttt{G} \equiv \wFold{F}{S}$, for some fixed $F$, and $S$.
Then, \texttt{G} is accepted as extension of $\bfAPI$ if:
(i) a combinator, say $\mathtt{K}\hasType\lType{A}\lImpl B$,
\emph{already} exists in the library, and
(ii) by induction on the length of every $l\hasType \lType{A}$, it holds that
$\lcA{\texttt{G}}{l}$ and $\lcA{\mathtt{K}}{l}$ compile to the same \lcLamTerm
in \DLAL, forcefully of type $B$.
For example, following \cite{DBLP:journals/jfp/Hutton99}, there are $H, F, S$
the such that
$\ppMapFromFold{H} \equiv \wFold{F[H]}{S}\hasType\lType{\bTypeC{2}}\lImpl
\lType{\bTypeC{2}}$ extends $\bfAPI$. In it $\bTypeC{2}$ is
the type of booleans, and $H\hasType \bTypeC{2}\lImpl \bTypeC{2}$, and $S$ the
empty list, and $F$ a wrapper such that $F[H]\hasType\bTypeC{2}\lImpl
\lType{\bTypeC{2}}\lImpl \lType{\bTypeC{2}}$. The extension exists because:
(i) $\ppMap{H}\hasType\lType{\bTypeC{2}}\lImpl \lType{\bTypeC{2}}$ belongs
to $\bfAPI$, and
(ii) both $\lcA{\ppMapFromFold{H}}{l}$, and $\lcA{\ppMap{H}}{l}$
compile to the same \lcLamTerm in \DLAL, for every $l$.
\par
$\wMT{\cdot}$ and $\wMS{\cdot}$ are further examples of programming patterns
that the $\bfAPI$ exposes, and that a programmer can exploit to extend the library,
under the above scheme. Section~\ref{section:Conclusion} will discuss ideas
about how extending and automatizing the use of our $\bfAPI$.
}
\subsubsection{Related Works on Polynomial Time Languages.}
A programming language inspired by Haskell is described in
\cite{DBLP:conf/esop/BaillotGM10}. The programs that can be developed in it belong to
the class of \polytime functions because the language inherits the principles
of the \lcLamTerms, or, equivalently, of the proof-nets of \LAL
\cite{Asperti+Roversi:2002-ToCL}. However, we are not aware of any attempt to
exploit it to program libraries with a real potential impact. The approach of
\cite{DBLP:conf/esop/BaillotGM10} to the development of a real programming
language for \polytime computations is quite orthogonal to ours. We proceed
bottom-up, showing that a reasonably interesting library can be developed inside
\DLAL. Then, we import standard programming patterns which were compatible with
the typing discipline of \DLAL. In \cite{DBLP:conf/esop/BaillotGM10}, the
language is given under the assumption that its primitives will really be used.

The same occurs in \cite{DBLP:conf/fossacs/BaillotM04} and \cite{Burrel:LCC09}.
The former extends \lcLamCalc to give formulas of \SLL
\cite{DBLP:journals/tcs/Lafont04}. The latter introduces $\mathsf{POLA}$, a
programming language which mixes object oriented and recursion schemes
for which an interpreter is also available\footnote{\url{http://projects.wizardlike.ca/projects/pola}}. The best developed
project we
are aware of, and which brings theoretical results related to the world of
\polytime bounded functions ``down to'' the practical level, is based on~\cite{DBLP:conf/popl/Hofmann02,DBLP:journals/iandc/Hofmann03}. The language
exploits formulas of a smartly crafted version of multiplicative linear logic as
types and is based on recursion schemes \`a la System T. We are still far from those
levels of migration of theory to practice.
\par
Our main distinguishing feature is to remain loyal to the theoretical
properties of \DLAL, while allowing programming with standard patterns of
functional programming.
\section{Typed Functional Assembly}

\subsubsection{\lcLamCalc.}
Given a set $\lcVar$, which we range over by \emph{any lowercase Latin letter},
the set $\lcSet$ of \lcLamTerms, which we range over by the \emph{uppercase
Latin letters} $M, N, P, Q, R$, this set contains terms generated as follows:
\begin{align}
\label{align:lambda-calculus-syntax}
M & \BNFDef \lcVar
       \mid \lcF{x}{M}
       \mid \lcA{M}{M}.
\end{align}
The set of free variables in $M$ is $\lcFV{M}$.
The set $\lcSetNF$ of \emph{values of our computations}, which we range over by
\emph{the uppercase
Latin letters} $V, W, X$, is defined as follows:
\begin{align}
\label{align:lambda-calculus-values}
V & \BNFDef \lcVar
       \mid \lcF{x}{V}
       \mid \lcA{x}{V}.
\end{align}
We remark that $\lcSetNF$ coincides the standard \emph{$\beta$ normal forms}.
\subsubsection{\BigSteps Rewriting Relation on \lcLamCalc.}
\begin{figure}
\fbox{
\begin{minipage}{.97\textwidth}
{\small
\[
\infer[\lcNFv]
      {\lcNFJud{x}{x}
      }{}
\qquad\quad
\infer[\lcNFf]
      {\lcNFJud{\lcF{x}{M}}
               {\lcF{x}{V}}
      }{\lcNFJud{M}
                {V}
       }
\qquad\quad
\infer[\lcNFava]
        {
         \lcNFJud{\lcA{M}{N}}
                 {\lcA{x}{V}}
        }{
          \lcNFJud{M}
                  {x}
          &
          \lcNFJud{N}
                  {V}
        }
\qquad\quad
\infer[\lcNFavl]
        {
         \lcNFJud{\lcA{M}{N}}
                 {W}
        }{
          \lcNFJud{M}
                  {\lcF{x}{V}}
          &
          \lcNFJud{V\lcSub{N}{x}}
                  {W}
        }
\]
}
\end{minipage}
}
\caption{Big steps rewriting relation $\lcNFJud{}{}$ on $\lcSet$ with results in $\lcSetVal$}
\label{fig:Big steps rewriting relation on lcSet with results in lcSetVal}
\end{figure}

The relation $\lcNFJud{\!}{\,}\subset \lcSet\!\times\!\lcSetNF$ is inductively defined in
Figure~\ref{fig:Big steps rewriting relation on lcSet with results in lcSetVal}.
\begin{figure}
\fbox{
\begin{minipage}{.97\textwidth}
{\small
\[
\infer[\taAx]
      {
       \taJudg{\taCntxEmpty}{\taCntxB{x}{A}}{x}{A}
      }{
}
\qquad\qquad
\infer[\taW]{
           \taJudg{\taCntxE,\taCntxE'}{\taCntxL,\taCntxL'}{M}{A}
        }{
           \taJudg{\taCntxE}{\taCntxL}{M}{A}
        }
\qquad\qquad
\infer[\taC]{
           \taJudg{\taCntxE,\taCntxB{z}{A}}
                  {\taCntxL}{M\lcSubT{z}{x}{z}{y}}{B}
        }{
          \taJudg{\taCntxE,\taCntxB{x}{A},\taCntxB{y}{A}}
                 {\taCntxL}{M}{B}
        }
\]
\[
\infer[\taLII]
        {
          \taJudg{\taCntxE}{\taCntxL}{\lcF{x}{M}}{A\lImpl B}
        }{
          \taJudg{\taCntxE}{\taCntxL,\taCntxB{x}{A}}{M}{B}
        }
\qquad\qquad
\infer[\taLIE]
        {
           \taJudg{\taCntxE,\taCntxE'}{\taCntxL,\taCntxL'}{\lcA{M}{N}}{B}
        }{
          \taJudg{\taCntxE}{\taCntxL}{M}{A\lImpl B}
          &
          \taJudg{\taCntxE'}{\taCntxL'}{N}{A}
        }
\]
\[
\infer[\taEII]{
          \taJudg{\taCntxE}
                 {\taCntxL}
                 {\lcF{x}{M}}
                 {\lBang A\lImpl B}
        }{
          \taJudg{\taCntxE
                 ,\taCntxB{x}{A}
                 }
                 {\taCntxL}
                 {M}{B}
        }
\qquad\qquad
\infer[\taEIE]{
               \taJudg{\taCntxE,\taCntxE'}
                      {\taCntxL}
                      {\lcA{M}{N}}{B}
              }{\taJudg{\taCntxE}
                       {\taCntxL}
                       {M}{\lBang A\lImpl B}
          &
          \taJudg{\taCntxEmpty}
                 {\taCntxE'}
                 {N}{A}
          &
          \Size{\taCntxE'}\leq 1
        }
\]
\[
\infer[\taPI]{
           \taJudg{\taCntxE}{\lPar\taCntxL}{M}{\lPar A}
        }{
           \taJudg{\taCntxEmpty}{\taCntxE, \taCntxL}{M}{A}
        }
\qquad\qquad
\infer[\taPE]{
           \taJudg{\taCntxE,\taCntxE'}{\taCntxL,\taCntxL'}{M\lcSub{N}{x}}{B}
        }{
           \taJudg{\taCntxE}{\taCntxL}{N}{\lPar A}
           &
          \taJudg{\taCntxE'}{\taCntxB{x}{\lPar A},\taCntxL'}{M}{B}
        }
\]
\[
\infer[\taQI]{
           \taJudg{\taCntxE}{\taCntxL}{M}{\forall \alpha.A}
        }{
           \taJudg{\taCntxE}{\taCntxL}{M}{A}
           &
           \alpha\not\in\taFV{\taCntxE,\taCntxL}
        }
\qquad\qquad\qquad
\infer[\taQE]{
           \taJudg{\taCntxE}{\taCntxL}{M}{A\taSub{B}{\alpha}}
        }{
           \taJudg{\taCntxE}{\taCntxL}{M}{\forall \alpha.A}
        }
\]
}
\end{minipage}
}
\caption{Type assignment system \TFA}
\label{fig:Type assignment system TFA}
\end{figure}

\subsection{Type assignment}
We introduce a type assignment \TFA\ which gives formulas of \LinearLogic\ as
types to \lcLamTerms. In fact, \TFA\ is \DLAL\ \cite{BaillotT09} whose set of formulas is
quotiented by a specific recursive equation. We recall that adding a recursive
equation among the formulas does not negatively affect polynomial time
soundness of \DLAL\ normalization which only depends on the structural
constraints that the process of formula construction puts on the form of
derivations \cite{GirardLLL}.

\subsubsection{Types for \TFA.}
Given a set $\taVar$ of \emph{formula variables}, which we range over by
\emph{lowercase Greek letters}, the set $\taFor$ of \emph{formulas}, that we
range over by the \emph{uppercase Latin letters} $A, B, C, D$, is defined as
follows:
\begin{align*}
A & \BNFDef \alpha
       \mid A \lImpl A
       \mid \lBang A \lImpl A
       \mid \forall \alpha. A
       \mid \lPar  A
\end{align*}

Note that   modal formulas $\lBang A$ can occur in negative positions only.
We obtain the set of \emph{types} $\taTyp$ when we consider the quotient of $\taFor$
 by the following fix-point equation:
{\small
\begin{equation}
 \Def{\sTypeC}
      {\forall\alpha. \sType{\alpha}}
\end{equation}
}
where   $\Def{\sType{\alpha}}
      {
        (\bTypeC{2} \lImpl \alpha) \lImpl
        ((\bTypeC{2}\lTime\sTypeC) \lImpl \alpha) \lImpl \alpha
     }$
and  $\bTypeC{2}$ is defined in Figure~\ref{fig:Relevant (defined) data}.
We say $\sTypeC$ is the type of \emph{Sequences}.
Thus, we actually use formulas which are equivalence classes of types in
$\taTyp$.
 
Note that once we use $\sTypeC$ as type of a \lcLamTerm\ $M$, we can equivalently
use any of its ``unfolded forms'' as type of $M$ as well.
In Figure~\ref{fig:Relevant (defined) data}, we also introduce relevant types we
use to develop our first level library.
As a notation, $A\taSub{B}{\alpha}$ is the clash free substitution of $B$
for every free occurrence of $\alpha$ in $A$ (here, clash-free means that occurrences of free variables of $B$ are
not bound in $A\taSub{B}{\alpha}$).
\begin{figure}
\begin{center}
{\small
\begin{tabular}{|c|c|}\hline
\textbf{Type}
& \textbf{Definitions}
\\\hline\hline
{
\begin{minipage}{.3\textwidth}
Finite types
\end{minipage}
}
&
{
\begin{minipage}{.5\textwidth}
\begin{align*}
& \Def{\bType{n}{\alpha}}
      {\overbrace{\alpha \lImpl \cdots \lImpl \alpha}^{n+1} \lImpl \alpha}
\\
& \Def{\bTypeC{n}}
      {\forall\alpha. \bType{n}{\alpha}}
\end{align*}
\end{minipage}
}
\\\hline
{
\begin{minipage}{.3\textwidth}
Tuples
\end{minipage}
}
&
{
\begin{minipage}{.5\textwidth}
\begin{align*}
& \Def{\pType{A_1}{\Seq{\lTime}{\lTime}}{A_n}{\alpha}}
      {A_1 \lImpl \cdots \lImpl A_n \lImpl \alpha}
\\
& \Def{\pTypeC{A_1}{\Seq{\lTime}{\lTime}}{A_n}}
      {\forall\alpha. \pType{A_1}{\Seq{\lTime}{\lTime}}{A_n}{\alpha}\lImpl\alpha}
\end{align*}
\end{minipage}
}
\\\hline
{
\begin{minipage}{.3\textwidth}
Church numerals
\end{minipage}
}
&
{
\begin{minipage}{.5\textwidth}
\begin{align*}
& \Def{\uType{\alpha}}
      {\lBang(\alpha \lImpl \alpha) \lImpl \lPar(\alpha \lImpl \alpha)}
\\
& \Def{\uTypeC}{\forall\alpha.\uType{\alpha}}
\end{align*}
\end{minipage}
}
\\\hline
{
\begin{minipage}{.3\textwidth}
Lists
\end{minipage}
}
&
{
\begin{minipage}{.5\textwidth}
\begin{align*}
& \Def{\lTypeG{A}{\alpha}}{\lBang(A \lImpl \alpha \lImpl \alpha)
\lImpl
\lPar(\alpha \lImpl \alpha)}
\\
& \Def{\lType{A}}{\forall\alpha. \wTypeG{A}{\alpha}}
\end{align*}
\end{minipage}
}
\\\hline
{
\begin{minipage}{.3\textwidth}
Church words
\end{minipage}
}
&
{
\begin{minipage}{.5\textwidth}
\begin{align*}
& \Def{\wTypeC{2}}{\lType{\bTypeC{2}}}
\end{align*}
\end{minipage}
}
\\\hline
\end{tabular}
}
\end{center}
\vspace*{-1em}
\caption{Relevant (defined) types}
\label{fig:Relevant (defined) data}
 \end{figure}
\subsubsection{Type assignment \TFA.} We give the type assignment system \TFA\ in Figure~\ref{fig:Type assignment system TFA}. In this formal system, 
we have  judgments of the form $\taJudg{\taCntxE}{\taCntxL}{M}{A}$ where
 \emph{context} $\taCntxE$ is \emph{exponential}, while context $\taCntxL$ is \emph{linear}.
Any context is a finite domain function $\Seq{x_1\hasType A_1,}{,x_n\hasType A_n}$ with domain
$\Set{\Seq{x_1,}{,x_n}}$, and range  $\Set{\Seq{A_1,}{,A_n}}$ in the codomain of the set of types.
Every pair $\taCntxB{x}{A}$  of any kind of context is a \emph{type assignment for a variable}.
\subsubsection{Tuples as primitives.}
The definition of tuples in Figure~\ref{fig:Relevant (defined) data} supports the introduction of
the tuples as primitives, as follows.
Extending \lcLamCalc\ with tuples means adding the
following clauses to~\eqref{align:lambda-calculus-syntax}:
\begin{align}
\label{align:lambda-calculus-syntax-with-pairs}
M & \BNFDef\ldots \mid \lcTC{\Seq{M,}{,M}} \mid \lcTD{\Seq{x,}{,x}}{M}.
\end{align}
So, values in~\eqref{align:lambda-calculus-values} also include:
\begin{align}
\label{align:lambda-calculus-values-with-pairs}
V & \BNFDef\ldots \mid \lcTC{\Seq{V,}{,V}},
\end{align}
and the set of rules in Figure~\ref{fig:Big steps rewriting relation on lcSet with results in
lcSetVal} must contain:
\[
\infer[\lcNFp]
      {\lcNFJud{\lcTC{\Seq{M_1,}{,M_n}}}
               {\lcTC{\Seq{V_1,}{,V_n}}}
      }{\lcNFJud{M_1}
                {V_1}
        &\ldots&
        \lcNFJud{M_n}
                {V_n}
       }
\]
\[
\infer[\lcNFap]
        {
         \lcNFJud{\lcA{M}{N}}
                 {W}
        }{
          \lcNFJud{M}
                  {\lcTD{\Seq{x_1,}{,x_n}}{V}}
          &
          \lcNFJud{N}
                  {\lcTC{\Seq{V_1,}{,V_n}}}
          &
          \lcNFJud{V\lcSubM{V_1}{x_1}{V_n}{x_n}}
                  {W}
        }
\]
Finally, we add the following derivable rules to those ones in Figure~\ref{fig:Type assignment
system TFA}:
\[
 \infer[\taTup]
        {
           \taJudg{\Seq{\taCntxE_1}{\taCntxE_n}}
                  {\Seq{\taCntxL_1}{\taCntxL_n}}
                  {\lcTC{\Seq{M_1,}{,M_n}}}
                  {\pTypeC{A_1}{\Seq{\lTime}{\lTime}}{A_n}}
        }{
          \taJudg{\taCntxE_1}
                 {\taCntxL_1}
                 {M_1}
                 {A_1}
          &
          \ldots
          &
          \taJudg{\taCntxE_n}
                 {\taCntxL_n}
                 {M_n}
                 {A_n}
        }
\]
\[
\infer[\taLIIp]
        {
          \taJudg{\taCntxE}
                 {\taCntxL}
                 {
                  \lcTD{\Seq{x_1,}{,x_n}}{M}
                  }
                 {
                 \pTypeC{A_1}{\Seq{\lTime}{\lTime}}{A_n}\lImpl B
                 }
        }{
          \taJudg{\taCntxE}
                 {\taCntxL
                 ,\Seq{\taCntxB{x_1}{A_1}}
                      {\taCntxB{x_n}{A_n}}
                 }
                 {M}{B}
        }
\]
Saying that the here above rules are derivable means that we use tuple as
abbreviations, as follows:
\begin{align}
\label{align:tuple-contructor-shortening}
\lcTC{\Seq{M_1,}{,M_n}} & \Def{}{}
     \lcF{x}{\lcA{\ldots(\lcA{x}{M_1})\ldots}{M_n}}\\
\label{align:tuple-destructor-shortening}
\lcTD{\Seq{x_1,}{,x_n}}{M}
                     & \Def{}{}
     \lcF{p}{\lcA{p}{\lcF{x_1}{\ldots\lcF{x_n}{M}}}}
\end{align}
%
\section{A Library for Binary Field Arithmetic}
\label{section:A Library for Binary Field Arithmetic}

In this section, we present a library of lambda-terms for the arithmetic in binary 
fields written in \DLAL.
The library is organized in functional layers, 
as shown in Figure~\ref{fig:Library for binary field arithmetic}.

The lowest layer contains basic definitions and it is interpreter-specific.
We have currently implemented the library with LCI\footnote{\url{http://lci.sourceforge.net}},
an interpreter for pure \lcLamCalc. 
We thus needed to define basic types, such as Church words, or 
\DLAL-specific combinators.
The \emph{core library} layer contains all the combinators to work on
basic types. We put particular care in the definition of common 
functional-programming patterns in \DLAL, and to reuse them,
whenever possible, while defining other combinators.
Finally, in the \emph{binary field arithmetic} layer we group all the
combinators related to operations over binary polynomials, like addition,
multiplication and modular reduction.

In future work, we plan to extend the library by implementing other layers,
such as arithmetic of elliptic curves or other cryptographic primitives, on 
top of the binary field arithmetic layer.

\vspace*{-1em}
\begin{figure}[h!]
	\centering
	\begin{tabularx}{0.7\linewidth}{|X|}
	\hline \emph{\textbf{Cryptographic primitives}: elliptic curves cryptography, \dots}\\
	\hline \textbf{Binary field arithmetic}: addition, (modular reduction), square, multiplication, \emph{inversion}.\\
	\hline \textbf{Core library}: operations on bits (xor, and),
	 operations on sequences (head-tail splitting), operations on words (reverse, drop, conversion to sequence, projections); meta-combinators: fold, map, mapthread, map with state.\\
	\hline \textbf{Basic definitions and types}: booleans, tuples, numerals, words, sequences, basic type management and duplication.\\
	\hline
	\end{tabularx}
	\vspace*{-.5em}
	\caption{Library for binary field arithmetic}
	\label{fig:Library for binary field arithmetic}
\end{figure}
\vspace*{-1em}

In the following subsections we present type and behaviour of the 
relevant combinators, while the full definition as \lcLamTerm is
in Appendix~\ref{section:Definition of Combinators}.
\subsection{Basic Definitions and Types}

\begin{figure}
	\begin{center}
	{\small
	\begin{tabular}{|c|c|}\hline
	& \textbf{(typed) Values}
	\\\hline\hline
	{
	\begin{minipage}{.2\textwidth}
	Booleans
	\end{minipage}
	}
	&
	{
	\begin{minipage}{.5\textwidth}
	\begin{align*}
	&\Def{\bTT }{\lcF{x y z}{x}}   \hasType \bTypeC{2}
	\\
	&\Def{\bFF }{\lcF{x y z}{y}}   \hasType \bTypeC{2}
	\\
	&\Def{\bBot}{\lcF{x y z}{z}}   \hasType \bTypeC{2}
	\end{align*}
	\end{minipage}
	}
	\\\hline
	{
	\begin{minipage}{.2\textwidth}
	Tuples
	\end{minipage}
	}
	&
	{
	\begin{minipage}{.5\textwidth}
	\begin{align*}
	&\Def{\pTupC{\Seq{M_1,}{,M_n}}}{\lcF{p}{\lcA{\ldots(\lcA{p}{M_1})\ldots}{M_n}}}
	 \hasType
	 \pTypeC{A_1}{\Seq{\lTime}{\lTime}}{A_n}
	\end{align*}
	\end{minipage}
	}
	\\\hline
	{
	\begin{minipage}{.2\textwidth}
	Church numerals
	\end{minipage}
	}
	&
	{
	\begin{minipage}{.5\textwidth}
	\begin{align*}
	& \Def{\uNil}{\lcF{f x}{x}} \hasType \uTypeC
	\\
	&\Def{\uExt{n}}
	     {\lcF{f x}{\underbrace{\lcA{f}{\dots\lcA{f}{}}}_n x}} \hasType \uTypeC
	\end{align*}
	\end{minipage}
	}
	\\\hline
	{
	\begin{minipage}{.2\textwidth}
	Church words
	\end{minipage}
	}
	&
	{
	\begin{minipage}{.5\textwidth}
	\begin{align*}
	& \Def{\wNil}{\lcF{f x}{x}} \hasType \wTypeC{2}
	\\
	& \Def{\wExt{\Seq{\bBit{n-1}}{\bBit{0}}}}
	      {\lcF{f x}{\lcA{\lcA{f}{\bBit{n-1}}}{\dots\lcA{\lcA{f}{\bBit{0}}}{x}}}}
	\hasType \wTypeC{2}
	\end{align*}
	\end{minipage}
	}
	\\\hline
	{
	\begin{minipage}{.2\textwidth}
	\Sequences
	\end{minipage}
	}
	&
	{
	\begin{minipage}{.5\textwidth}
	\begin{align*}
	& \Def{\sNil}
	      {\lcF{t c}{\lcA{t}{\bBot}}} \hasType \sTypeC
	\\
	& \Def{\sExt{\Seq{\bBit{n-1}}{\bBit{0}}}}
	      {\lcF{t c}{
		         \lcA{c}
		             {
		              \lcPC{\bBit{n-1}}
		                   {\sExt{\Seq{\bBit{n-2}}{\bBit{0}}}}
		             }
		        }
	      }
	\hasType \sTypeC
	\end{align*}
	\end{minipage}
	}
	\\\hline
	\end{tabular}
	}
	\end{center}
	\vspace*{-1em}
	\caption{Canonical values of data-types}
	\label{fig:Relevant types and canonical values}
\end{figure}

In Figure~\ref{fig:Relevant types and canonical values}, we give names to those formulas which are \emph{types} we actually use in the library and we identify the
\lcLamTerms\ that we define as \emph{canonical values} of the corresponding type.
In every \Sequence\ $\sExt{\Seq{\bBit{n-1}}{\bBit{0}}}$ and Church word
$\wExt{\Seq{\bBit{n-1}}{\bBit{0}}}$
the \emph{least significant bit} (\blsb) is $\bBit{0}$, while the \emph{most
significant bit} (\bmsb) is $\bBit{n-1}$.

\noindent
In \DLAL, we can derive the rule \emph{paragraph lift}:
\[
\infer[\taPL]
        {
          \taJudg{\taCntxEmpty}
                 {\taCntxEmpty}
                 {
                  \lcPL{}{M}
                 }
                 {\lPar A\lImpl\lPar B}
        }{
          \taJudg{\taCntxEmpty}
                 {\taCntxEmpty}
                 {M}{A\lImpl B}
        }
\]
where $\Def{\lcPL{}{M}}{\lcF{x}{\lcA{M}{x}}}$ is the \emph{paragraph lift of
$M$}. As obvious generalization, $n$ consecutive applications of the $\taPL$ rule define
a lifted term
$\Def{\lcPL{n}{M}} {\lcF{x}{\lcA{\ldots\lcF{x}{\lcA{M}{x}}\ldots}{x}}}$, that
contains $n$ nested $\lcPL{}{\cdot}$. Its type is $\lPar ^n A\lImpl\lPar^n B$.
Borrowing terminology from proof nets, the application of $n$ paragraph lift of
$M$ \emph{embeds} it in $n$ paragraph boxes, leaving the behaviour of $M$
unchanged:
\begin{align*}
\lcNFJud{\lcA{\lcPL{n}{M}}{N}}
        {\lcA{M}{N}}.
\end{align*}

\par\noindent
The combinator $\bCast{m}\hasType\bTypeC{2}\lImpl\lPar^{m+1}\bTypeC{2}$ embeds a boolean into
$m+1$ paragraph boxes, without altering the boolean:
\begin{align*}
\lcNFJud{\lcA{\bCast{m}}{\bBit{}}}{\bBit{}}.
\end{align*}
\par\noindent
The combinator $\bDup{t}\hasType
\bTypeC{2}\lImpl(\overbrace{\bTypeC{2}\lTime\cdots\lTime\bTypeC{2}}^{t})$, for
every $t\geq 2$, produces $t$ copies of a boolean:
\begin{align*}
\lcNFJud{\lcA{\bDup{t}}
             {\bBit{}}
        }
        {\lcTC{\overbrace{\Seq{\bBit{},}
                              {,\bBit{}}
                         }^t
              }.
        }
\end{align*}

\par\noindent
The combinator
$\tCast{m}\hasType
(\bTypeC{2}\lTime\bTypeC{2})\lImpl\lPar^{m+1}(\bTypeC{2}\lTime\bTypeC{2})$, for every
$m\geq0$, embeds a pair of bits into $m+1$ paragraph boxes, without altering the
structure of the pair:
\begin{align*}
\lcNFJud{\lcA{\tCast{m}}{\lcPC{\bBit{0}}{\bBit{1}}}}
        {\lcPC{\bBit{0}}{\bBit{1}}}.
\end{align*}

\par\noindent
The combinator 
$\wSuc \hasType \bTypeC{2} \lImpl \wTypeC{2} \lImpl \wTypeC{2}$ implements the
\emph{successor} on Church words:
\begin{align*}
&
\lcNFJud{
	\lcA{\lcA{\wSuc}
	         {\bBit{}}
	    }
	    {\wExt{\Seq{\bBit{n-1}}{\bBit{0}}}
	    }
         }
  {\wExt{\bBit{}\ \Seq{\bBit{n-1}}{\bBit{0}}}}.
\end{align*}

\par\noindent
The combinator $\wCast{m}\hasType\wTypeC{2}\lImpl\lPar^{m+1}\wTypeC{2}$, for every $m\geq0$,
embeds a word into $m+1$ paragraph boxes, without altering the structure of the word:
\begin{align*}
\lcNFJud{\lcA{\wCast{m}}{\wExt{\Seq{\bBit{n-1}}{\bBit{0}}}}}
        {\wExt{\Seq{\bBit{n-1}}{\bBit{0}}}}.
\end{align*}
\par\noindent
The combinator $\wDup{t}{m}\hasType
\wTypeC{2}\lImpl\lPar^{m+1}(\overbrace{\wTypeC{2}\lTime\cdots\lTime\wTypeC{2}} ^{t})$, for
every $t\geq2$, $m\geq0$, produces $t$ copies of a word deepening the result
into $m+1$ paragraph boxes:
\begin{align*}
\lcNFJud{\lcA{\wDup{t}{m}}
             {\wExt{\Seq{\bBit{n-1}}{\bBit{0}}}}
        }
        {\lcTC{\overbrace{\Seq{\wExt{\Seq{\bBit{n-1}}{\bBit{0}}},}
                              {,\wExt{\Seq{\bBit{n-1}}{\bBit{0}}}}
                         }^t
              }
        }.
\end{align*}

\subsection{Core Library}

\subsubsection{Operations on Bits.}\
\par\noindent
The combinator $\bXor\hasType \bTypeC{2}\lImpl \bTypeC{2}\lImpl \bTypeC{2}$ extends
the \emph{exclusive or} as follows:
\begin{align*}
\lcNFJud{((\bXor) \bFF ) \bFF }{\bFF }& &
\lcNFJud{((\bXor) \bTT ) \bTT }{\bFF } \\
\lcNFJud{((\bXor) \bFF ) \bTT }{\bTT }& &
\lcNFJud{((\bXor) \bTT ) \bFF }{\bTT } \\
\lcNFJud{((\bXor) \bBot) \bBit{}}{\bBit{}}& &
\lcNFJud{((\bXor) \bBit{}) \bBot}{\bBit{}}& & \mbox{(where $\bBit{} \hasType \bTypeC{2}$)}. 
\end{align*}
Whenever one argument is $\bBot$ then it gives back the other argument.
This is an application oriented choice. Later we shall see why.
\par\noindent
The combinator $\bAnd\hasType \bTypeC{2}\lImpl \bTypeC{2}\lImpl \bTypeC{2}$ extends the
\emph{and} as follows:
\begin{align*}
\lcNFJud{((\bAnd) \bFF ) \bFF }{\bFF }& &
\lcNFJud{((\bAnd) \bTT ) \bTT }{\bTT } \\
\lcNFJud{((\bAnd) \bFF ) \bTT }{\bFF }& &
\lcNFJud{((\bAnd) \bTT ) \bFF }{\bFF } \\
\lcNFJud{((\bAnd) \bBot) \bBit{}}{\bBot}& &
\lcNFJud{((\bAnd) \bBit{}) \bBot}{\bBot}& & \mbox{(where $\bBit{} \hasType \bTypeC{2}$)}. 
\end{align*}
Whenever one argument is $\bBot$ then the result is $\bBot$.
Again, this is an application oriented choice.

\subsubsection{Operations on \Sequences.}\
\par\noindent
The combinator $\sSplit\hasType \sTypeC\lImpl(\bTypeC{2}\lTime\sTypeC)$
\emph{splits} the \sequence it takes as input in a pair with the \bmsb and the
corresponding tail:
\begin{align*}
\lcNFJud{\lcA{\sSplit}{\sExt{\Seq{\bBit{n-1}}{\bBit{0}}}}}
         {\lcPC{\bBit{n-1}}{\sExt{\Seq{\bBit{n-2}}{\bBit{0}}}}}.
\end{align*}

\subsubsection{Operations on Church Words.}\
\par\noindent
The combinator $\wRev \hasType \wTypeC{2} \lImpl \wTypeC{2}$
\emph{reverses the bits} of a word:
\begin{align*}
\lcNFJud{\lcA{\wRev}{\wExt{\Seq{\bBit{n-1}}{\bBit{0}}}}}
        {\wExt{\Seq{\bBit{0}}{\bBit{n-1}}}}.
\end{align*}
\par\noindent
The combinator $\wDropB \hasType \wTypeC{2} \lImpl \wTypeC{2}$
\emph{drops} all the (initial) occurrences\footnote{%
The current definition actually drops all the occurrences of $\bBot$ in
a Church word, however we shall only apply $\wDropB$ to words that contain
$\bBot$ in the most significant bits.
} of $\bBot$ in a word:
\begin{align*}
\lcNFJud{\lcA{\wDropB}{\wExt{\Seq{\bBot}{\bBot}\ \Seq{\bBit{n-1}}{\bBit{0}}}}}
        {\wExt{\Seq{\bBit{n-1}}{\bBit{0}}}}.
\end{align*}
\par\noindent
The combinator
$\wTos \hasType \wTypeC{2}\lImpl\lPar\sTypeC$
\emph{casts} a word into a \sequence:
\begin{align*}
\lcNFJud{\lcA{\wTos}{\wExt{\Seq{\bBit{n-1}}{\bBit{0}}}}}
        {\sExt{\Seq{\bBit{n-1}}{\bBit{0}}}}.
\end{align*}
\par\noindent
The combinator
$\wProj\hasType \lType{\bTypeC{2}^2}\lImpl\wTypeC{2}$ \emph{projects} the first
component of a list of pairs:
\begin{align*}
&
\lcNFJud{\lcA{\wProj}
             {\lcF{f x}
                  {\lcA{\lcA{f}{\lcPC{\bBita{n-1}}{\bBit{n-1}}}}
                       \ldots
                       {\lcA{\lcA{f}{\lcPC{\bBita{0}}{\bBit{0}}}}{x}}
		  }
             }
         }
         {\wExt{\Seq{\bBita{n-1}}{\bBita{0}}}}
\enspace.
\end{align*}
Similarly, $\wProjb \hasType \lType{\bTypeC{2}^2}\lImpl\wTypeC{2}$ projects the second
component. The argument of $\wProj$ has not the form
$\wExt{\Seq{\lcPC{\bBita{n-1}}{\bBit{n-1}}}
           {\lcPC{\bBita{0}}{\bBit{0}}}}$ because its elements are not booleans. We
shall adopt the same convention also for the forthcoming meta-combinators.

\subsubsection{Meta-combinators on Lists.}
Meta-combinators are \lcLamTerms with one or two ``holes'' that allow to use
standard higher-order programming patterns to extend the $\bfAPI$. Holes must be
filled with type constrained \lcLamTerms.  We discuss how to
use meta-combinators in order to effectively implement arithmetic in
Section~\ref{binaryfields}, after their introduction here below.
\medskip
\par\noindent
The first meta-combinator we deal with is $\wMap{\cdot}$.
Let $\wMapPar\hasType A\lImpl B$ be a closed term. 
Then, $\wMap{\wMapPar} \hasType \lType{A}\lImpl\lType{B}$ applies
$\wMapPar$ to every element of the list that $\wMap{\wMapPar}$ takes as argument,
and yields the final list, assuming $\lcNFJud{\lcA{\wMapPar}{\bBit{i}}}{\bBitP{i}} $,
for every $0\leq i\leq n-1$:
\begin{align*}
& \lcNFJud{\lcA{\wMap{\wMapPar}}
               {\lcF{f x}
                    {\lcA{\lcA{f}{\bBit{n-1}}}
                         \ldots
                         {\lcA{\lcA{f}{\bBit{0}}}{x}}
		    }
               }}
          {\lcF{f x}{\lcA{\lcA{f}{\bBitP{n-1}}}
                         \ldots
                       {\lcA{\lcA{f}{\bBitP{0}}}{x}}}}
\end{align*}
\par\noindent
The second meta-combinator is $\wFold{\cdot}{\cdot}$.
Let $\wMapPar\hasType A \lImpl B \lImpl B$ and $\wMapParS\hasType B$ be closed
terms. Let also $\Cast{0}\hasType B \lImpl \lPar B$.
Then, $\wFold{\wMapPar}{\wMapParS} \hasType \lType{A}\lImpl \lPar B$,
starting from the initial value $\wMapParS$,
iterates $\wMapPar$ over the input list and builds up a value, assuming 
$\lcNFJud{\lcA{\lcA{\wMapPar}{\bBit{i}}}{\bBitP{i}}}{\bBitP{i+1}} $, for
every $0\leq i\leq n-1$, and setting $\bBitP{0} = \wMapParS$ and $\bBitP{n} =
\bBitP{}$:
\begin{align*}
& \lcNFJud{\lcA{\wFold{\wMapPar}{\wMapParS}}
               {\lcF{f x}
                    {\lcA{\lcA{f}{\bBit{n-1}}}
                         \ldots
                         {\lcA{\lcA{f}{\bBit{0}}}{x}}
		    }
	       }}
          {\bBitP{}}          
\end{align*}
\par\noindent
The third meta-combinator is $\wMS{\cdot}$.
Let $\wMapPar \hasType (A \lTime S) \lImpl (B \lTime S)$ be a closed term.
Then, $\wMS{\wMapPar} \hasType \lType{A} \lImpl S \lImpl \lType{B}$ applies
$\wMapPar$ to the elements of the input list, keeping track of a
\emph{state} of type $S$ during the iteration. Specifically,
if
$\lcNFJud{\lcA{\wMapPar}{\lcPC{\bBit{i}}{\bBits{i}}}}{\lcPC{\bBitP{i}}{\bBits{i+1}}}
$, for every $0\leq i\leq n-1$:
\begin{align*}
& \lcNFJud{\lcA{\lcA{\wMS{\wMapPar}}
                    {
                     \lcF{f x}
                         {\lcA{\lcA{f}{\bBit{n-1}}}
                             \ldots
                              {\lcA{\lcA{f}{\bBit{0}}}{x}}
		         }
		    }
               }
               {\bBits{0}}
           }
           {
                \lcF{f x}
                    {\lcA{\lcA{f}{\bBitP{n-1}}}
                         \ldots
                         {\lcA{\lcA{f}{\bBitP{0}}}{x}}
		    }
	   }
\end{align*}
\par\noindent
Finally, the fourth meta-combinator is $\wMT{\cdot}$.
Let $\wMapPar\hasType \bTypeC{2} \lImpl \bTypeC{2} \lImpl A$ be a closed term. 
Then, $\wMT{\wMapPar} \hasType \wTypeC{2}\lImpl\wTypeC{2}\lImpl\lType{A}$ applies
$\wMapPar$ to the elements of the input list. Specifically, if
$\lcNFJud{\lcA{\lcA{\wMapPar}{\bBita{i}}}{\bBit{i}}}{\bBitc{i}} $, for every
$0\leq i\leq n-1$:
\begin{align*}
& \lcNFJud{\lcA{\lcA{\wMT{\wMapPar}}{\wExt{\Seq{\bBita{n-1}}{\bBita{0}}}}}{\wExt{\Seq{\bBit{n-1}}{\bBit{0}}}}}
          {\lcF{f x}
               {\lcA{\lcA{f}{\bBitc{n-1}}}
                    \ldots
                    {\lcA{\lcA{f}{\bBitc{0}}}{x}}
               }
          }
\end{align*}
\noindent
In particular, $\wMT{\lcF{a}{\lcF{b}{\lcPC{a}{b}}}}\hasType
\wTypeC{2}\lImpl\wTypeC{2}\lImpl\lType{\bTypeC{2}^2}$ is such that:
\begin{align*}
&
\lcNFJud{\lcA{\lcA{\wMT{\lcF{a}{\lcF{b}{\lcPC{a}{b}}}}}
                  {\wExt{\Seq{\bBita{n-1}}{\bBita{0}}}}
             }{\wExt{\Seq{\bBit{n-1}}{\bBit{0}}}}
        }
\\&\qquad\qquad\qquad\qquad\qquad
        { \lcF{f x}
              {\lcA{\lcA{f}{\lcPC{\bBita{n-1}}{\bBit{n-1}}}}
                    \ldots
                   {\lcA{\lcA{f}{\lcPC{\bBita{0}}{\bBit{0}}}}{x}}
              }
        }
\end{align*}

\ifodd2
\subsection{Deleted}
The combinator $\uSuc\hasType\uTypeC\lImpl\uTypeC$ gives the successor of a Church numeral:
\begin{align*}
\lcNFJud{\lcA{\uSuc}
             {\uExt{n}}
        }
        {\uExt{n+1}}
\end{align*}
\par\noindent
The combinator $\uCast{m}\hasType\uTypeC\lImpl\lPar^{m+1}\uTypeC$ embeds a Church numeral into
$m+1$ paragraph boxes, without altering the structure of the numeral:
\begin{align*}
\lcNFJud{\lcA{\uCast{m}}{\uExt{n}}}{\uExt{n}}
\end{align*}
The combinator $\uPred\hasType\uTypeC\lImpl\uTypeC$ gives the predecessor of a Church numeral:
\begin{align*}
\lcNFJud{\lcA{\uPred}
             {\uExt{n}}
        }
        {\uExt{n-1}}
\end{align*}
\par\noindent
The combinator $\uDup{m}{n}\hasType
\uTypeC\lImpl\lPar^{n+1}(\overbrace{\uTypeC\lTime\cdots\lTime\uTypeC} ^{m})$ produces $m$
copies of a Church numeral deepening the result into $n+1$ paragraph boxes:
\begin{align*}
\lcNFJud{\lcA{\uDup{m}{n}}
             {\uExt{p}}
        }
        {\lcTC{\overbrace{\Seq{\uExt{p}}
                              {\uExt{p}}
                         }^m
              }
        }
\end{align*}
\par\noindent
The combinator $\uDiff\hasType\uTypeC\lImpl\uTypeC\lImpl\lPar\uTypeC$ gives the difference of
two Church numerals:
\begin{align*}
&\lcNFJud{\lcA{\lcA{\uDiff}
                   {\uExt{m}}}{\uExt{n}}
         }
         {\uExt{m-n}}
 \textrm{ if } m > n
\\
& \lcNFJud{\lcA{\lcA{\uDiff}
                   {\uExt{m}}}{\uExt{n}}
          }
          {\uExt{0}}
\qquad \textrm{ otherwise}
\end{align*}

\par\noindent
The combinator $\wTou\hasType\wTypeC{2}\lImpl\uTypeC$ converts a word into a Church numeral,
giving the legth of the word:
\begin{align*}
\lcNFJud{\lcA{\wTou}
             {\wExt{\Seq{\bBit{0}}{\bBit{p-1}}}}
        }
        {\uExt{p}}
\end{align*}
\par\noindent
The combinator $\wLenghtDiff\hasType\wTypeC{2}\lImpl\wTypeC{2}\lImpl\lPar\uTypeC$ gives the
length difference of two words:
\begin{align*}
&
\lcNFJud{\lcA{\lcA{\wLenghtDiff}
                  {\wExt{\Seq{\bBit{0}}{\bBit{p-1}}}}
             }
             {\wExt{\Seq{\bBitP{0}}{\bBitP{q-1}}}}
        }
        {\uExt{p-q}}
\textrm{ if } p > q
\\
&
\lcNFJud{\lcA{\lcA{\wLenghtDiff}
                  {\wExt{\Seq{\bBit{0}}{\bBit{p-1}}}}
             }
             {\wExt{\Seq{\bBitP{0}}{\bBitP{q-1}}}}
        }
        {\uExt{0}}
\qquad\textrm{ otherwise }
\end{align*}

\par\noindent
The combinators $\wSucZ, \wSucO \hasType \wTypeC{2} \lImpl \wTypeC{2}$ implement the
\emph{successor} on words by adding a \blsb\ to their argument:
\begin{align*}
&
\lcNFJud{\lcA{\wSucZ}{ \wExt{\Seq{\bBit{0}}{\bBit{n-1}}}}}
        {\wExt{\bFF,\Seq{\bBit{0}}{\bBit{n-1}}}}
\\
&
\lcNFJud{\lcA{\wSucO}{ \wExt{\Seq{\bBit{0}}{\bBit{n-1}}}}}
        {\wExt{\bTT,\Seq{\bBit{0}}{\bBit{n-1}}}}
\end{align*}
\par\noindent
The combinator $\wShiftRBot\hasType\uTypeC\lImpl\wTypeC{2}\lImpl\wTypeC{2}$, for every $n> 0$,
adds a certain number of $\bBot$ as \blsb\ to a word:
\begin{align*}
\lcNFJud{\lcA{\lcA{\wShiftRBot}
                  {\uExt{n}}
             }
             {\wExt{\Seq{\bBit{0}}{\bBit{m-1}}}}
        }
        {\wExt{\underbrace{\Seq{\bBot}{\bBot}}_n\Seq{\bBit{0}}{\bBit{m-1}}}}
\end{align*}
It implies
$\Def{\wSucBot}
     {\lcF{n}{\lcA{\lcA{\wShiftRBot}{\uExt{1}}}{n}}}$.
\par\noindent
The combinator $\wInitBot \hasType \wTypeC{2} \lImpl \wTypeC{2}$, \emph{initializes} a word
with all $\bBot$ elements in it:
\begin{align*}
\lcNFJud{\lcA{\wInitBot}{\wExt{\Seq{\bBit{0}}{\bBit{n-1}}}}}
        {\wExt{\Seq{\bBot}{\bBot}}}
\end{align*}

\subsubsection{Map thread $\wMT$ on lists}
The combinator $\wMT\hasType \lType{A}\lImpl\lType{B}\lImpl\lType{A\lTime B}$
produces a list of pairs that
couple elements with the same position in the two arguments lists,
the \emph{proviso} being $n\geq m$:
\begin{align*}
\lcNFJud{
         \lcA{\lcA{\wMT}
                  {\wExt{\Seq{\bBitR{0}}{\bBitR{n-1}}}}}
             {\wExt{\Seq{\bBitg{0}}{\bBitg{m-1}}}}
        }
        {
         \wExt{\Seq{\lcPC{\bBitR{0}}{\bBot}}
                   {\lcPC{\bBitR{n-m-1}}{\bBot}}
               \Seq{\lcPC{\bBitR{n-m}}{\bBitg{0}}}
                   {\lcPC{\bBitR{n-1}}{\bBitg{m-1}}}
              }
        }
\end{align*}
\subsubsection{Map unthread $\wMU$ on lists}
The combinator $\wMU\hasType\lType{A\lTime B}\lImpl\lType{A}$ assumes the
second component of every element of the input list is garbage. So, it erases
every of them an gives the list with the first elements:
\begin{align*}
\lcNFJud{\lcA{\wMU}
             {\wExt{\Seq{\lcPC{\bBit{0}  }{\bBitP{0}}}
                        {\lcPC{\bBit{m-1}}{\bBitP{m-1}}}
                   }
             }
        }
        {\wExt{\Seq{\bBit{0}  }
                   {\bBit{m-1}}
              }
        }
\end{align*}

\subsubsection{Map state $\wMS{\cdot}$ on lists}
We here introduce the meta-combinator $\wMS{\cdot}$.
Let $\wMapPar\hasType A\lImpl B\lImpl C\lImpl ( D\lTime
C)$ be a closed term,
and the type $C$ such that $\Cast{0}\hasType C\lImpl\lPar C$
exists. Also, let
$\wMapParB\hasType \alpha\lTime C\lImpl\alpha$ be a closed term. The combinator
$\wMS{\wMapParB,\wMapPar}\hasType C\lImpl\lType{A\lTime B}\lImpl \lType{D}$
produces a list with elements of type $D$ that we obtain applying $\wMapPar$ to
the pairs of the input and to the state of type $C$:
\begin{align*}
\lcNFJud{\lcA{\lcA{\wMS{\wMapParB,\wMapPar}}
                  { s_0
                  }
             }
             {
               \wExt{\Seq{\lcPC{\bBitR{0}  }{\bBitg{0}}}
                             {\lcPC{\bBitR{m-1}}{\bBitg{m-1}}}
                        }
             }
        }
        {\wExt{\Seq{\bBitRP{0}  }
                   {\bBitRP{m-1}}
              }
        }
\end{align*}
if $\lcNFJud{\lcA{\lcA{\lcA{\wMapPar}{\bBitR{i}}}{\bBitg{i}}}{s_{i}}}
{\lcPC{\bBitRP{i}}{s_{i+1}}} $, for every $0\leq i\leq m-1$. The reason why $\wMS{\cdot}$ is a
meta-combinator is the same as for $\wMap{\cdot}$.
\subsubsection{Bit-wise xor $\wXorBW$ on words}
The combinator $\wXorBW \hasType \wTypeC{2}\lImpl\wTypeC{2}\lImpl\wTypeC{2}$
applies $\bXor$ to
every pair of elements with the
same position in the two input lists. The result is a word:
\begin{align*}
\lcNFJud{\lcA{\lcA{\wXorBW}
                  {\wExt{\Seq{\bBit{0}}{\bBit{n-1}}}}
             }
             {\wExt{\Seq{\bBitP{0}}{\bBitP{n-1}}}}
        }
        {\wExt{\Seq{\bBitPP{1}}{\bBitPP{n-1}}}}
\end{align*}
where $\lcNFJud{((\bXor) \bBit{i}) \bBitP{i}}{\bBitPP{i}}$, for every $0\leq
i\leq n-1$.
\fi
\subsection{Binary Field Arithmetic}\label{binaryfields}

We start by recalling the essentials on
binary field arithmetic. For wider details we address the reader to~\cite[Section
11.2]{avanzi05handbook}.
Let $p(X) \in \F_2[X]$ be an
irreducible polynomial of degree $n$ over $\F_2$,
and let $\beta \in \overline{\F}_2$ be a root of $p(X)$ in the algebraic closure
of $\F_2$.
Then, the finite field $\F_{2^n} \simeq \F_2[X] / (p(X)) \simeq \F_2(\beta)$. 

The set of elements $\{1,\beta,\ldots,\beta^{n-1}\}$ is a basis of $\F_{2^n}$ 
as a vector space over $\F_{2}$ and we can represent a generic element 
of $\F_{2^n}$ as a polynomial in $\beta$ of degree lower than $n$:
$$
   \F_{2^n} \ni a = \sum_{i=0}^{n-1}a_i\beta^i=a_{n-1} \beta^{n-1}
              + \dotsb + a_1 \beta + a_0
   \enspace,\qquad
   a_i \in \F_2
   \enspace.
$$
Moreover, the isomorphism $\F_{2^n} \simeq \F_2[X] / (p(X))$ allows us to
implement the arithmetic of $\F_{2^n}$ relying on the arithmetic of $\F_2[X]$
and reduction modulo $p(X)$.

Since each element $a_i \in \F_2$ can be encoded as a bit, we can represent each
element of $\F_{2^n}$ as a Church word of bits of type $\wTypeC{2}$.

In what follows, we denote by $\wModParN$ the Church numeral representing
$n = \deg p(X)$, and by $\wModParF$ the Church word: 
$\Def{\wModParF}{\wExt{\Seq{p_{n}}{p_0}\underbrace{\Seq{\bBot}{\bBot}}_{n-1}}}\enspace,$
where $p_i$ are such that $p(X) = \sum p_i X^i$. 
Note that $\wModParF$ has length $2n-1$. The $\bBot$ in the least significative part
are included for technical reasons, to simplify the discussion later.

\subsubsection{Addition.}
Let $a, b \in \F_{2^n}$.
The addition $a+b$ is computed component-wise, i.e., setting $a = \sum a_i \beta^i$
and $b = \sum b_i \beta^i$,
then $a+b = \sum (a_i+b_i) \beta^i$. 
The sum $(a_i+b_i)$ is done in $\F_2$ and corresponds to the bitwise exclusive or.
This led us to the following definition:
\par\noindent
The combinator
$\bfAdd\hasType\bfType\lImpl\bfType\lImpl\bfType$ is:
\begin{equation}
\Def{\bfAdd}{
	\wMT{\bXor}
}
\end{equation}

\subsubsection{Modular Reduction.}
Reduction modulo $p(X)$ is a fundamental building block to keep the size of the operands constrained.
We implemented a na{\"i}f left-to-right method, assuming that:
(1) both $p(X)$ and $n=\deg p(X)$ are fixed (thus axioms); 
(2) the length of the input is $2n$, i.e. we need exactly $n$ repetitions of a basic iteration.
\par\noindent
The combinator
$\wMod{\wModParN}{\wModParF}\hasType\wTypeC{2}\lImpl\lPar\bfType$ is:
\begin{align*}
\Def{\wMod{\wModParN}{\wModParF}}{&\
\lcF{d}{
	\lcA{\lcPL{}{\wModEnd}}\\&\
	    {\lcA{\lcA{\wModParN}
	              {\lcF{l}
	                   {\lcA{\lcA{\wMS{\wModFun}}
	                             {l}
	                        }
	                        {\lcPC{\bBot}{\bFF}}
	                   }
	              }
	         }
                 {\lcA{\wModBase[\wModParF]}{\lcA{\wCast{0}}{d}}}
	    }
	}
}
\end{align*}
where: 
\begin{align*}
\Def{\wModEnd}{&\ \lcF{l}{\lcA{\wDropB}{\lcA{\wRev}{\lcA{\wProj}{l}}}}}\\
\Def{\wModFun}{&\
	\lcPD{e}{s}{
	\lcA{\lcPD{d}{p}{
	\lcA{\lcPD{s_0}{s_1}{
	\\&\ \ \lcA{
	       \lcA{\lcA{\lcA{\lcA{\lcA{\lcA{s_0}{		
		    \lcF{d p s}{				
				 \lcA{\lcPD{p'}{p''}
				          {\lcPC{\lcPC{\lcA{\lcA{\bXor}{d}}{p'}}{s}}
                                                {\lcPC{\bTT}{p''}}
                                          }
                                     }
				     {\lcA{\bDup{2}}{p}}
			        }
	                                         }
		                       }
		                  }{\\&\qquad\ \ \ \ \ \ \ \
		   \lcF{d p s}
		       {\lcPC{ \lcPC{d}{s} }{ \lcPC{\bFF}{p} }}
		                   }
		           }{\\&\qquad\ \ \ \ \ \ \ \
		               \lcF{d p s}
		               {\lcPC{\lcPC{\bBot}{s}}
		                     {\lcPC{d}    {p }}
		               } 
                            }
                }{d}
            }{p}
         }{s_1}	
\\&\	}}{s}}
	}{e}
	}
}\\
\Def{\wModBase[\wModParF]}{&\
	\lcF{d}{
	\lcA{\lcA{\wMT{\lcF{a}{\lcF{b}{\lcPC{a}{b}}}}}{
		\lcA{\wRev}{d}
	}}{
		\lcA{\wRev}{\wModParF}
	}}
}
\end{align*}
The basic iteration is implemented via $\wMS{\cdot}$, that operates on a list of bit
pairs
$\wExt{\ldots \lcPC{\bBitd{i}}{\bBitp{i}} \ldots}$, where $\bBitd{i}$ are the bits of the input and $\bBitp{i}$ the bits of $\wModParF$.
The core of the algorithm is the combinator 
$\wModFun \hasType (\bTypeC{2}^2\,\lTime\,\bTypeC{2}^2) \lImpl (\bTypeC{2}^2\,\lTime\,\bTypeC{2}^2)$,
that behaves as follows:
\begin{align*}
\lcNFJud{
	\lcA{\lcA{\wModFun}{
		\underbrace{\lcPC{\bBitd{i}}{\bBitp{i}}}_{\text{elem. }e}
	}}{
		\underbrace{\lcPC{\bBits{0}}{\bBitp{i+1}}}_{\text{status }s}
	}
}{
	\lcPC{
		\underbrace{\lcPC{\bBitd{i}'}{\bBitp{i+1}}}_{e'}
	}{
		\underbrace{\lcPC{\bBits{0}'}{\bBitp{i}}}_{s'}
	}
}\enspace,
\end{align*}
where $\bBits{0}$ keeps the \bmsb of $\wExt{\ldots \bBitd{i} \ldots}$ and 
it is used to decide wether to reduce or not at this iteration.
Thus, $\bBitd{i}' = \bBitd{i} + \bBitp{i}$ if $\bBits{0} = \bTT$;
$\bBitd{i}' = \bBitd{i}$ if $\bBits{0} = \bFF$; 
and $\bBitd{i}' = \bBot$ when $\bBits{0} = \bBot$ 
(that represents the initial state, when $\bBits{0}$ still needs to be set).

Note that the second component of the status is used to shift $\wModParF$
(right shift as the words have been reverted).

\subsubsection{Square.}
Square in binary fields is a linear map (it is the absolute Frobenius automorphism).
If $a \in \F_{2^n},\ a = \sum a_i \beta^i$, then $a^2 = \sum a_i \beta^{2i}$.
This operation is obtained by inserting zeros between the bits that represent $a$
and leads to a polynomial of degree $2n-2$, that needs to be reduced 
modulo $p(X)$.

Therefore, we introduce two combinators: $\wSqr\hasType\wTypeC{2}\lImpl\wTypeC{2}$
that performs the bit expansion, and
$\bfSqr\hasType\bfType\lImpl\lPar\bfType$ that is the actual square in $\F_{2^n}$.
We have:
\begin{equation}
\Def{\bfSqr}{
	\lcF{a}{
		\lcA{\wMod{\wModParN}{\wModParF}}{\lcA{\wSqr}{a}}
	}
}
\end{equation}
and
$\Def{\wSqr}{
\lcF{lfx}{
	\lcA{\lcA{l}{
		\wSqrStep[f]
	}}{x}
}
}$, where
$\Def{\wSqrStep[f]}{
	\lcF{et}{
		\lcA{\lcA{f}{
			\bFF
		}}{
			\lcA{\lcA{f}{
				e
			}}{t}
		}
	}
}$ has type $\bTypeC{2}\lImpl\alpha\lImpl\alpha$ if $f$ is a non linear variable
with type $\bTypeC{2}\lImpl\alpha\lImpl\alpha$.

\subsubsection{Multiplication.}
Let $a, b \in \F_{2^n}$.
The multiplication $ab$ is computed as polynomial multiplication,
i.e., with the usual definition,
$ab = \sum_{j+k=i} (a_j+b_k) \beta^i$. 
\par
We currently implemented the na{\"i}ve schoolbook method. A possible extension to the
\emph{comb method} is left as future straightforward work. On the contrary, it is not clear how to implement
the Karatsuba algorithm, which reduces the multiplication of $n$-bit words
to operations on $n/2$-bit words. The difficulty is to represent the
splitting of a word in its half upper and lower parts.
\par
Similarly as for the square, we have to distinguish between the polynomial multiplication
$\wMult\hasType\wTypeC{2}\lImpl\wTypeC{2}\lImpl\lPar\wTypeC{2}$
and the field operation $\bfMult\hasType\bfType\lImpl\bfType\lImpl\lPar^2\bfType$,
obtained by composing with the modular reduction.
We have:
\begin{align*}
\bfMult
& \Def{}{}
	\lcF{ab}{
		\lcA{\lcPL{}{\wMod{\wModParN}{\wModParF}}}{
			\lcA{\lcA{\wMult}{a}}{b}
		}
	}
\\
\wMult
& \Def{}{}
\lcF{a b}
     {\lcA{\lcPL{}{\wProjb}}
          {\lcA{\lcA{b}
                    {\lcF{M l}
                         {\lcA{\lcA{\wMultStep}
                                   {\lcPC{M}{\bBot}}
                              }
                              {l}
                         }
                    }
               }
               {\lcA{\wMultBase}{\lcA{\wCast{0}}{a}}}
          }
    }
\end{align*}
\par
The internals of $\wMult$ are in Figure~\ref{fig:Multiplication: definition of the
combinators}. It implements two nested iterations. The parameter $b$ controls
the external, and $a$ the internal one.
\begin{figure}[t]
	\centering
	{\small
	\begin{tabular}{|c|}\hline
	{
	\begin{minipage}{.97\textwidth}
		\begin{align*}
		\Def{\wMultStep}
		{&\
		\lcF{s l f x}
		     {\lcA{\wBMult{f}}
			  {\lcA{\lcA{l}{\wMSStep{f,\wFMult}}
			       }
			       {\lcA{\wMSBase{x}}
				    {\lcA{\tCast{0}}{s}
				    }
			       }
			  }
		     }
		}
		\\
		\Def{\wMultBase}
		{&\
		\lcF{m}{
			\lcA{\lcA{\wMT{\lcF{a}{\lcF{b}{\lcPC{a}{b}}}}}{m}}{\wNil}
		}}
		\\
		\Def{\wMSStep{f,\wFMult}}
		{&\
		\lcF{e}
		     {
		      \lcPD{w}{s}
			   {
			    \lcA{
				  \lcPD{e'}{s'}
				       {
				        \lcPC{\lcA{\lcA{f}{e'}}
				                  {w}
				             }
				             {s'}
				       }
				}
				{
				  \lcA{\lcA{\wFMult}{e}}{s}
				}
			   }
		    }
		}
		\\
		\Def{\wMSBase{x}}
		{&\ 
			\lcF{s}{\lcPC{x}{s}}
		}
		\\
		\Def{\wFMult}
		{
		&\ \lcPD{m}{r}
		     {\lcPD{M}{\bar m}
			   {\lcA{\lcPD{m'}{m''}
				      {
					\lcA{\lcPD{M'}{M''}
				                 {\\&\qquad
				                    \lcPC{\lcPC{\bar m}
				                             {\lcA{\lcA{\bXor}
				                                       {\lcA{\lcA{\bAnd}
				                                                 {m'}
				                                            }
				                                            {M'}
				                                        }
				                                  }
				                                  {r}
				                              }
				                       }
				                       {\lcPC{M''}{m''}
						       }
				                 }\\&\
				            }
				            {\lcA{\bDup{2}}{m}}
				      }
				 }{\lcA{\bDup{2}}{M}}
			     }
		      }
		}
		\\
		\Def{\wBMult{f}}
		{&\
		\lcPD{w}{s}
		      {\lcA{\lcPD{M}{\bar m}
				 {
				  \lcA{\lcA{f}
				           {\lcPC{\bar m}{\bFF}}
				      }
				  {w}
				 }
			   }
			   {s}
		      }
		}
		\\
		\end{align*}
	\end{minipage}
	}
	\\\hline
	\end{tabular}
	}
	\vspace*{-.5em}
	\caption{Multiplication: definition of the combinators}
	\label{fig:Multiplication: definition of the combinators}
\end{figure}

\par
The external iteration (controlled by $b$) works on words of bit pairs.
The combinator $\wMultStep\hasType
\bTypeC{2}^2 \lImpl \lType{\bTypeC{2}^2} \lImpl \lType{\bTypeC{2}^2}$ behaves as follows:
\begin{align*}
\lcNFJud{
	\lcA{\lcA{\wMultStep}{
		\lcPC{\bBitM{}}{\bBot}
	}}{
           \lcF{f x}
               {\ldots
                \lcA{\lcA{f}
                         {\lcPC{\bBitm{i}}{\bBitr{i}}}
                    }
                    {\ldots x}
               }
	}
}{
           \lcF{f x}
               {\ldots
                \lcA{\lcA{f}
                         {\lcPC{\bBitm{i-1}}{\bBitrP{i}}}
                    }
                    {\ldots x}
               }
}
\end{align*}
where $\bBitM{}$ is the current bit of the Multiplier $b$, and every $\bBitm{i}$ is
a bit of the multiplicand $a$, and every $\bBitr{i}$ is a bit in the current result.
The iteration is enabled by
the combinator $\wMultBase\hasType
\wTypeC{2} \lImpl \lType{\bTypeC{2}^2}$,
that, on input $a$, creates
$\lcF{f x}
    {\lcA{\lcA{f}{\lcPC{\bBitm{n-1}}{\bBot}}}
         \ldots
         {\lcA{\lcA{f}{\lcPC{\bBitm{0}}{\bBot}}}{x}}
    }$, setting the initial bits of the result to $\bBot$.
The projection $\wProjb$ returns the result when the iteration stops.
\par
The internal iteration is used to update the above list of bit pairs.
The core of this iteration is 
the combinator $\wFMult\hasType
\bTypeC{2}^2 \lImpl \bTypeC{2}^2 \lImpl (\bTypeC{2}^2\,\lTime\,\bTypeC{2}^2)$,
that behaves as follows:
\begin{align*}
\lcNFJud{
	\lcA{\lcA{\wFMult}{
		\underbrace{\lcPC{\bBitm{i}}{\bBitr{i}}}_{\text{elem. }e}
	}}{
		\underbrace{\lcPC{\bBitM{}}{\bBitm{i-1}}}_{\text{status }s}
	}
}{
	\lcPC{
		\underbrace{\lcPC{\bBitm{i-1}}{\bBitM{}\cdot\bBitm{i}+\bBitr{i}}}_{e'}
	}{
		\underbrace{\lcPC{\bBitM{}}{\bBitm{i}}}_{s'}
	}
}\enspace.
\end{align*}
For completeness, we list the type of the other combinators:
$
\wMSStep{f,\wFMult} \hasType
\bTypeC{2}^2 \lImpl (\alpha\,\lTime\,\bTypeC{2}^2) \lImpl (\alpha\,\lTime\, \bTypeC{2}^2) \enspace,\enspace
\wMSBase{x} \hasType
\bTypeC{2}^2 \lImpl (\alpha\,\lTime\,\bTypeC{2}^2) \enspace,\enspace
\wBMult{f} \hasType
(\alpha\,\lTime\,\bTypeC{2}^2) \lImpl \alpha \enspace.
$

\subsubsection{Inversion.}
It is under development. We are concentrating on the binary Euclidean algorithm,
which is the ``left-to-right'' counterpart of the extended Euclidean algorithm (for a
detailed analysis, we refer to Fong et al.~\cite{fong04field}).%
\section{Developing (with) the Library} \label{sec:developing}

Beside the implementation of the library, we experimented the use of
higher-order combinators to improve the readability of the code, as
well as the programming experience.
Inspired by~\cite{DBLP:journals/jfp/Hutton99}, we have rewritten some 
combinators relying on standard programming pattern such as $\wMap{\cdot}$
and $\wFold{\cdot}{\cdot}$, ``simulating'' the behavior of a programmer
that wants to add new functionality to the library.
The idea is to let the programmer write a combinator in a more comfortable style, 
and then to \emph{compile} the combinator to a value that admits a type in \DLAL.
In the following, we give some relevant examples of increasing difficulty.

We know that $\wTos$ is defined as
$\Def{\wTos}{
\lcF{l}
 {
  \lcA{
       \lcA{l}
           {
            \lcF{e s t c}{\lcA{c}{\lcPC{e}{s}}}
           }
     }
     {\sNil}
}}$.
A programmer could anyway define it by using the
standard programming pattern $\wFold{\cdot}{\cdot}$ as follows:
$$\Def{\wTosFromFold}{
\wFold{
	\lcF{e s t c}{
		\lcA{c}{\lcPC{e}{s}}
	}
}
{\sNil}
}$$
The combinator $\wTosFromFold$ is a legal one because
$\wTosFromFold$ compiles exactly to $\wTos$. The compilation consists of in-line
substituting the parameters of $\wFold{\cdot}{\cdot}$ and of applying the
rewriting steps in Figure~\ref{fig:Big steps rewriting relation on lcSet with
results in lcSetVal}, whose key intermediate \lcLamTerms are
$\lcF{l}{
	\lcA{\lcA{l}{						
		\lcF{ez}{					
			\lcA{					
				\lcF{t c}{
					\lcA{c}{\lcPC{e}{z}}
				}
			}{					
				e
			}}
		}
	}}{							
		\sNil
	}$ and
$\lcF{l}{
  \lcA{
       \lcA{l}
           {
            \lcF{e z t c}{\lcA{c}{\lcPC{e}{z}}}
           }
     }
     {\sNil}
}$.

As a second example, we consider the combinator 
$\Def{\wProj}{
\lcF{lfx}{
	\lcA{\lcA{l}{						
		\lcPD{a}{b}{					
			\lcA{f}{				
				a				
			}
		}
	}}{x}
}}$, we define the following combinator and we show that
it is equivalent to the above one:
$$
\Def{\wProjFromMap}{\wMap{\lcPD{a}{b}{a}}}
$$
We recall that $\wProj \hasType \lType{\bTypeC{2}^2}\lImpl\wTypeC{2}$. 
While compiling the expression, we need
the assumption that each element $e$ of the input word is $\lcPC{a'}{b'} \hasType \bTypeC{2}^2$.
The key step is the reduction from
$\lcF{lfx}{
	\lcA{\lcA{l}{						
		\lcF{e}{					
			\lcA{f}{				
				\lcA{\lcPD{a}{b}{a}}{e}		
			}
		}
	}}{x}
}$ to
$\lcF{lfx}{
	\lcA{\lcA{l}{
		\lcPD{a'}{b'}{
			\lcA{f}{
				\lcA{\lcPD{a}{b}{a}}{\lcPC{a'}{b'}}
			}
		}
	}}{x}
}$\,, by replacing $\lcPC{a'}{b'}$ for $e$ in accordance with the assumption.

Finally, we show that the combinator
$\Def{\wMap{\wMapPar}}{
\lcF{lfx}{
	\lcA{\lcA{l}{						
		\lcF{e}{					
			\lcA{f}{				
				\lcA{\wMapPar}{e}		
			}
		}
	}}{x}
}}$
can be written using $\wFold{\cdot}{\cdot}$ 
(see also~\cite[Section 2]{DBLP:journals/jfp/Hutton99}) as:
$$\Def{\wMapFromFold{\wMapPar}}{
\wFold{
	\lcF{e p f x}{
		\lcA{\lcA{f}{
			\lcA{\wMapPar}{e}
		}}{
			\lcA{\lcA{p}{
				f
			}}{x}
		}
	}
}
{\wNil}
}$$
Here, the compilation process shows that $\lcA{\wMap{\wMapPar}}{l'}$ and
$\lcA{\wMapFromFold{\wMapPar}}{l'}$
are equivalent to the same value. We proceed by induction on the length of the
Church word $l'$.
First, we note that:
$$
\lcNFJud{\wMapFromFold{\wMapPar}}{
\lcF{l}{
	\lcA{\lcA{l}{						
		\lcF{e p f x}{
			\lcA{\lcA{f}{
				\lcA{\wMapPar}{e}
			}}{
				\lcA{\lcA{p}{
					f
				}}{x}
			}
		}
	}}{							
		\wNil
	}
}}$$
The base case is easy to check: $\lcNFJud{ \lcA{\wMap{\wMapPar}}{\wNil} }{ \wNil }$
and $\lcNFJud{ \lcA{\wMapFromFold{\wMapPar}}{\wNil} }{ \wNil }$.

We now prove the inductive case.
Let $\Def{l}{{\lcF{f x}{\lcA{\lcA{f}{\bBit{n-1}}}{\dots\lcA{\lcA{f}{\bBit{0}}}{x}}}}}$
be a Church word of length $n$.
Assume that $\lcNFJud{ \lcA{\wMap{\wMapPar}}{l} }{ V }$
and $\lcNFJud{ \lcA{\wMapFromFold{\wMapPar}}{l} }{ V }$.
We want to show that $\wMap{\wMapPar}$ and $\wMapFromFold{\wMapPar}$
reduce to the same term for a Church word 
$\Def{l'}{\lcF{fx}{ \lcA{\lcA{f}{\bBit{}}}  \lcA{\lcA{l}{f}} x}}$ of length $n+1$.
We report the key intermediate \lcLamTerms:
\begin{align}
\nonumber
\lcNFJud{\lcA{\wMap{\wMapPar}}{l'}}{
\ &
\lcF{fx}{
	\lcA{\lcA{l'}{
		\lcF{e}{
			\lcA{f}{
				\lcA{\wMapPar}{e}
			}
		}
	}}{x}
}}
\\
\label{align:Map-reduction-00}
\lcNFJud{}{\ &
\lcF{fx}{
		\lcA{
			\lcA{f}{
				\lcA{\wMapPar}{\bBit{}}
			}
		}
		\lcA{\lcA{l}{
			\lcF{e}{
				\lcA{f}{
					\lcA{\wMapPar}{e}
				}
			}
		}} x
}}
\\
\label{align:Map-reduction-01}
\lcNFJud{\scriptsize\text{ind. hyp. }}{\ &
\lcF{fx}{
		\lcA{
			\lcA{f}{
				\lcA{\wMapPar}{\bBit{}}
			}
		}
		\lcA{\lcA{V}{f}}{x}
}}
\\
\nonumber
\lcNFJud{\lcA{\wMapFromFold{\wMapPar}}{l'}}{
\ &
\lcA{\lcA{l'}{
	\lcF{e p f x}{
		\lcA{\lcA{f}{
			\lcA{\wMapPar}{e}
		}}{
			\lcA{\lcA{p}{
				f
			}}{x}
		}
	}
}}{							
	\wNil
}}
\\
\nonumber
\lcNFJud{}{\ &
	\lcA{\lcA{
		\lcF{e p f x}{
			\lcA{\lcA{f}{
				\lcA{\wMapPar}{e}
			}}{
				\lcA{\lcA{p}{
					f
				}}{x}
			}
		}
	}{\bBit{}}}  \lcA{\lcA{l}{
		\lcF{e p f x}{
			\lcA{\lcA{f}{
				\lcA{\wMapPar}{e}
			}}{
				\lcA{\lcA{p}{
					f
				}}{x}
			}
		}
	}} {\wNil}
}
\\
\nonumber
\lcNFJud{\scriptsize\text{ind. hyp. }}{\ &
	\lcA{\lcA{
		\lcF{e p f x}{
			\lcA{\lcA{f}{
				\lcA{\wMapPar}{e}
			}}{
				\lcA{\lcA{p}{
					f
				}}{x}
			}
		}
	}{\bBit{}}}
		V
}
\\
\nonumber
\lcNFJud{}{\ &
	\lcF{f x}{
		\lcA{\lcA{f}{
			\lcA{\wMapPar}{\bBit{}}
		}}{
			\lcA{\lcA{V}{
				f
			}}{x}
		}
	}}
\end{align}
This example is particularly relevant because
$\wMapFromFold{\wMapPar}\hasType\lType{A}\lImpl \lPar \lType{B}$, and
$\wMap{\wMapPar}\hasType \lType{A}\lImpl \lType{B}$ compile to a common term
despite their types differ. This is possible by applying two $\beta$-expansions
from \eqref{align:Map-reduction-00} to \eqref{align:Map-reduction-01}  which do
not duplicate any structure. 
\section{Conclusion and Future Work}
\label{section:Conclusion}
We have presented a core library for binary field arithmetic developed \DLAL.
The main motivation behind this work is to achieve a programming framework with
\emph{built-in} polynomial complexity and, from this perspective, 
this library is just a starting point, as
it lacks inversion and a complete realistic applicative example, such as elliptic curves cryptography. 
In the same line, the
implementation of symmetric-key cryptographic algorithms (block/stream ciphers,
hash functions,~\ldots) looks attractive as well, thanks to the higher-order
bitwise operations at the core of the current $\bfAPI$.
\par
Next, we shall investigate a full compilation process whose target will be
machine code. Namely, we plan to go further beyond the first compilation phase
of Section~\ref{sec:developing}, where, in fact, we describe an in-line
parameters unfolding of standard programming patterns like
$\wMap{\cdot}$ and $\wFold{\cdot}{\cdot}$. The compilation to machine code will target
parallelization, generally implied by functional programming thanks to its
reduced data dependency.
\par
\remove{
Another way of taking advantage of the functional paradigm is a specific form of
static code analysis, hard to achieve with common programming languages like C.
An example of what we mean is giving a strategy to prove that the square of
fields of characteristic 2 is linear, namely that $(x+y)^2 = x^2 + y^2$ holds.
Let us assume $(x+y)^2$ is represented by $\lcF{x}{\lcF{y}{M}}$, and $x^2 + y^2$
by $\lcF{x}{\lcF{y}{N}}$, Then, we can prove that the equivalence holds by
showing by induction on $x$, and $y$, that $\lcF{x}{\lcF{y}{M}}$, and
$\lcF{x}{\lcF{y}{N}}$ reduce to the same normal form. The large number of
projects that have sprung up to address problems analogous to the one just
illustrated witnesses the relevance of such kind
of program verification. Just as examples, we cite C Intermediate
Language\footnote{\url{http://www.cs.berkeley.edu/\~{}necula/cil}}, a full
infrastructure for C program analysis and transformation, and the Applied Type
System\footnote{\url{http://www.ats-lang.org}} (a programming language that
unifies specification and implementation.
\par
Then, \DLAL can be used for code certification, specifically to guarantee
polynomial time complexity, in scenarios like cloud computing where a client
want to send some code to a cloud platform and let the cloud perform the
computation on behalf of the client himself. Here we do not consider client-side
issues (privacy of his data, integrity of the computation and the result
\ldots), while we focus on guaranteeing that the client can not misbehave by,
e.g., flooding the cloud with too much computation, thus realizing a denial of
service attack.} Interestingly, while programming the binary field arithmetic, we
found that the main programming patterns we used can be assimilated to the
\textsf{MapReduce} paradigm~\cite{dean08mapreduce}. This means that not only
\DLAL can be used to certify polynomial-time complexity, but it is also suitable
to be adapted to actual cloud platforms based on the \textsf{MapReduce}.
\par
Finally, we do not exclude that more refined logics than \DLAL can be used to
realize a similar framework with even better built-in properties. Our choice of
\DLAL originated as a trade-off between flexibility in programming and
constrains imposed by the typing system, but it is at the same time an
experiment. Different logics can for instance measure the space complexity, or
provide a more fine-grained time complexity.


\bibliographystyle{splncs}
\bibliography{10LLF-DLAL-based.bib}

\appendix
\section{Definition of Combinators}
\label{section:Definition of Combinators}

\begin{description}
\item[$\bCast{m}$] is
$\lcF{b}
     {\lcA{\lcA{\lcA{b}
                 {\bTT}
            }{\bFF}
       }{\bBot}
     }$.
\item[$\bDup{t}$] is
$\lcF{b}
     {\lcA{\lcA{\lcA{b}
                 {\lcTC{\overbrace{\Seq{\bTT}{\bTT}}^t}}
            }
            {\lcTC{\overbrace{\Seq{\bFF}{\bFF}}^t}
            }
       }
       {\lcTC{\overbrace{\Seq{\bBot}{\bBot}}^t}
       }
     }$,
for every $t\geq 2$.
\item[$\tCast{m}$] is, for every $m\geq0$:
\begin{align*}
\tCast{0} &\equiv
\lcPD{a}{b}
 {\lcA{\lcA{\lcA{\lcA{a}
                     {\lcF{x}{\lcA{\lcA{\lcA{x}
                                            {\lcPC{\bTT}{\bTT }}
                                        }
                                        {\lcPC{\bTT}{\bFF }}
                                   }
                                   {\lcPC{\bTT}{\bBot}}
                              }
                      }
                 }
                 {\\ &\phantom{\qquad\qquad\qquad\ \ }
                     \lcF{x}{\lcA{\lcA{\lcA{x}
                                           {\lcPC{\bFF}{\bTT }}
                                      }
                                      {\lcPC{\bFF}{\bFF }}
                                 }
                                 {\lcPC{\bFF}{\bBot}}
                              }
                }
            }
           {\\ &\phantom{\qquad\qquad\qquad\ \ }
                     \lcF{x}{\lcA{\lcA{\lcA{x}
                                           {\lcPC{\bBot}{\bTT }}
                                      }
                                      {\lcPC{\bBot}{\bFF }}
                                 }
                                 {\lcPC{\bBot}{\bBot}}
                              }}
        }
        {b}
 }
\\
\tCast{m+1}&\equiv
{\lcF{p}
     {\lcA{\lcPL{}{\tCast{m}}}
          {\lcA{\tCast{0}}{p}}
     }
}
\enspace .
\end{align*}
\item[$\wSuc$] is
$
{\lcF{bp}{\lcF{f x}{\lcA{\lcA{f}{\lcA{\bCast{0}}{b}}}{\lcA{\lcA{p}{f}}{x}}}}}
$.
\item[$\wCast{m}$] is, for every $m\geq0$:
\begin{align*}
\wCast{0} &\equiv
{\lcF{l}
     {\lcA{\lcA{\lcA{\lcA{l}
                    {\lcA{\wSuc}{\bFF}}
               }
               {\lcA{\wSuc}{\bTT}}
               }
               {\lcA{\wSuc}{\bBot}}
          }
          {\wNil}
    }
}
\\
\wCast{m+1}&\equiv
{\lcF{l}
     {\lcA{\lcPL{}{\wCast{m}}}
          {\lcA{\wCast{0}}{l}}
     }
}
\enspace .
\end{align*}
\item[$\wDup{t}{m}$,]
for every $t \geq 2$, and $m \geq 0$ is:
\begin{align*}
\wDup{t}{0}
&
\Def{}{}
\lcF{l}
    {\lcA{\lcA{\lcA{l}
                   {\lcA{\wDupStep}{\bFF}
                   }
              }
              {\lcA{\wDupStep}{\bTT}
              }
         }
         {\wDupBase}
    }
\\
\wDup{t}{m+1}
&
\Def{}{}
\lcF{l}
    {\lcA{\lcPL{}
               {\wDup{t}{m}
               }
         }
         {\lcA{\wDup{t}{0}
              }
              {l}
         }
    }
\\
\wDupStep
&
\Def{}{}
\lcF{\bBit{}}{\lcTD{\Seq{x_1}{x_{t}}}{\lcTC{\overbrace{\Seq{
	\lcA{\lcA{\wSuc}{\bBit{}}}{x_1}
}{
	\lcA{\lcA{\wSuc}{\bBit{}}}{x_t}
}}^t}}}
\\
\wDupBase
&
\Def{}{}
\lcTC{\overbrace{\Seq{\wNil}{\wNil}}^t}
\enspace .
\end{align*}
\item[$\bXor$] is:
$\lcF{b c }
 {
  \lcA{
       \lcA{
            \lcA{
                 \lcA{b}
                     {\lcF{x}
                          {\lcA{\lcA{\lcA{x}{\bFF}}{\bTT}}{\bTT}}}
                }
                {\lcF{x}
                     {\lcA{\lcA{\lcA{x}{\bTT}}{\bFF}}{\bFF}}}
           }
           {\lcF{x}{x}}
      }
      {c}
 }$.
\item[$\bAnd$] is
$\lcF{bc}{
	\lcA{\lcA{\lcA{\lcA{b}{
		\lcF{x}{x}
	}}{
		\lcF{x}{
			\lcA{\lcA{\lcA{x}{
				\bFF
			}}{
				\bFF
			}}{
				\bBot
			}
		}
	}}{
		\bBot
	}}{c}
}$.
\item[$\sSplit$] is
$\lcF{s}
 {\lcA{
       \lcA{s}
           {\lcF{t}
                {\lcPC{\bBot}
                      {\sNil}
                }
           }
      }
      {\lcF{x}{x}}
 }$.
\item[$\wRev$] is
$\lcF{lfx}{
	\lcA{\lcA{\lcA{l}{
		\wRevStep{f}
	}}{
		\lcId
	}}{	x
	}
}$
with:\\
$\Def{\wRevStep{f}}
{
 \lcF{erx}
     {\lcA{r}
          {\lcA{\lcA{f}
                    {e}
               }
               {x}
          }
     }
}
\hasType
\bTypeC{2} \lImpl (\alpha\lImpl\alpha)\lImpl \alpha\lImpl\alpha$, when
$\taCntxB{f}{\bTypeC{2}\lImpl\alpha\lImpl\alpha}$.
\item[$\wDropB$] is
$\lcF{lfx}{
	\lcA{\lcA{l}{						
		\lcF{e}{					
			\lcA{\lcA{\lcA{\lcA{e}{			
				\lcF{f}{			
					\lcA{f}{\bTT}
				}
			}}{
				\lcF{f}{			
					\lcA{f}{\bFF}
				}
			}}{
				\lcF{fz}{z}			
			}}{f}					
		}
	}}{x}
}$.
\item[$\wTos$] is
$\lcF{l}
 {
  \lcA{
       \lcA{l}
           {
            \lcF{e s t c}{\lcA{c}{\lcPC{e}{s}}}
           }
     }
     {\sNil}
}$.
\item[$\wProj$] is
$\lcF{lfx}{
	\lcA{\lcA{l}{
		\lcPD{a}{b}{
			\lcA{f}{
				a
			}
		}
	}}{x}
}$.
\item[$\wProjb$] is
$\lcF{lfx}{
	\lcA{\lcA{l}{
		\lcPD{a}{b}{
			\lcA{f}{
				b
			}
		}
	}}{x}
}$.
\item[$\wMap{\wMapPar}$] is
$\lcF{lfx}{
	\lcA{\lcA{l}{						
		\lcF{e}{					
			\lcA{f}{				
				\lcA{\wMapPar}{e}		
			}
		}
	}}{x}
}$,
with $\wMapPar\hasType A\lImpl B$ closed.
\item[$\wFold{\wMapPar}{\wMapParS}$] is
$\lcF{l}{
	\lcA{\lcA{l}{						
		\lcF{ez}{					
			\lcA{\lcA{\wMapPar}{			
				e
			}}{z}
		}
	}}{							
		\lcA{\Cast{0}}{\wMapParS}			
	}
}$,
with $\wMapPar\hasType A \lImpl B \lImpl B$, and $\wMapParS\hasType B$ closed.
\item[$\wMS{\wMapPar}$] is
$\lcF{l s f x}{
	\lcA{\lcPD{w}{s'}{w}}{
		\lcA{\lcA{l}{
			\wMSStep{\wMapPar,f}
		}}{
			\lcA{\wMSBase{x}}{
				\lcA{\Cast{0}}{s}
			}
               }}
}$,
with\\
$\wMapPar\hasType (A\lTime S)\lImpl (B \lTime S)$ closed, and:
\begin{align*}
\wMSStep{\wMapPar,f}
& \Def{}{}
\lcF{e}{
	\lcPD{w}{s}{
		\lcA{\lcPD{e'}{s'}{
			\lcPC{\lcA{\lcA{f}{
				e'
			}}{
				w
			}}{
				s'
			}}
		}{
			\lcA{\wMapPar}{
				\lcPC{e}{s}
			}
                }
           }
    }
\hasType (A \lTime S) \lImpl (\alpha \lTime S)\lImpl (\alpha\lTime S)
\\
\wMSBase{x}
& \Def{}{} \lcF{s}{\lcPC{x}{s}}\hasType S\lImpl (\alpha \lTime S)
\enspace .
\end{align*}
\item[$\wMT{\wMapPar}$] is
$\lcF{l m f x}{
	\lcA{\lcPD{w}{s}{
		w
	}}{\lcA{\lcA{l}{
		\wMTStep{\wMapPar,f}
	}}{
		\lcA{\wMTBase{x}}{\lcA{\wTos}{\lcA{\wRev}{m}}}
	}}
}$,
with\\
$\wMapPar\hasType A \lImpl B \lImpl C$ closed,
$ \lcA{\wTos}{\lcA{\wRev}{m}} \hasType \lPar\sTypeC$, whenever
$m\hasType \wTypeC{2}$, and:
\begin{align*}
\wMTStep{\wMapPar,f}
&
\Def{}{}
\lcF{a}{\lcPD{w}{s}{
	\lcA{\lcPD{b}{s'}{
		\lcPC{
			\lcA{\lcA{f}{
				\lcA{\lcA{\wMapPar}{a}}{b}
			}}{w}
		}{s'}
	}}{
		\lcA{\sSplit}{s}
	}}
}\hasType
\bTypeC{2}\lImpl(\alpha\lTime\sTypeC)\lImpl(\alpha\lTime\sTypeC)
\\
\wMTBase{x}
&
\Def{}{}
\lcF{x}{\lcPC{x}{m}}
\hasType
\alpha\lImpl\alpha\lTime\sTypeC
\enspace .
\end{align*}
\end{description}

\ifodd2
\subsection{Deleted}
\begin{description}
\item[$\uSuc$] is
$\lcF{n f x}
          {\lcA{f}
               {\lcA{\lcA{n}
                         {f}
                    }
                    {x}
               }
     }
$.
\item[$\uCast{m}$] is
$\Def{\uCast{m}}
    {\lcF{l}
         {\lcA{l}
              {\lcPL{m}{\uSuc}}
         }
         {\wNil}
    }$,
for any $m\geq0$.
\item[$\uPred$] is
$\lcF{n f x}
     {\lcA{\lcPD{l}{r}
                   {r}}
          {\lcA{\lcA{n}
                    {\uPredStep{f}}
               }
               {\uPredBase{x}}
          }
     }$
with:
\begin{align*}
\uPredStep{f} & \Def{}{} \lcPD{g}{t}{\lcPC{f}{\lcA{g}{t}}}\\
\uPredBase{x} & \Def{}{} \lcPC{\lcF{z}{z}}{x}
\end{align*}
\item[$\uDup{m}{n}$] is
$\lcF{l}
         {\lcA{\lcA{l}
                   {\lcPL{n}
                          {\uDupStep}
                   }
              }
              {\uDupBase}
         }$,
for every $m, n\geq 0$, with:
\begin{align*}
& \Def{\uDupStep}
      {\lcTD{\Seq{x_1}{x_{n}}}{\lcTC{\overbrace{\Seq{\uSuc}{\uSuc}}^m}}}
\\
& \Def{\uDupBase}
      {\lcTC{\overbrace{\Seq{\uNil}{\uNil}}^m}}
\end{align*}
\item[$\uDiff$] is
$\lcF{m n}
          {\lcA{\lcA{n}{\uPred}
               }
               {\lcA{\uCast{0}}{m}}
          }$.
\item[$\wTou$] is
$\lcF{l}
     {\lcA{\lcA{l}
               {
		\lcF{etuc}{\lcA{c}
		               {\lcPC{e}{t}
		               }
		          }
               }
         }
         {\wNil}
    }$.
\item[$\wLenghtDiff$] is
$\lcF{l m}
     {\lcA{\lcA{\uDiff}
               {\lcA{\wTou}
	            {l}
               }
          }
          {\lcA{\wTou}
	       {m}
          }
     }$.
\item[$\wShiftRBot$] is
$\lcF{n l f x}
     {\lcA{\lcA{n}
               {\wShiftRBotStep{f}}
          }
          {\lcA{\lcA{l}
                    {f}
               }
               {x}
          }
      }$,
for every $n> 0$, where:
\begin{align*}
\Def{\wShiftRBotStep{f}}
    {\lcF{x}
         {\lcA{\lcA{f}
                   {\bBot}
              }
              {x}
         }
    }
\end{align*}
\item[$\wSucBot$] is $\lcF{n}{\lcA{\lcA{\wShiftRBot}{\uExt{1}}}{n}}$.
\item[$\wInitBot$] is
$\lcF{lfx}
     {\lcA{\lcA{l}
               {\lcBL{\lcF{x y}
                          {\lcA{\lcA{f}{\bBot}}{y}}
                     }
               }
          }
     }$.
\item[$\wSucZ$] is
$ \lcF{p}
 {
  \lcF{f x}
 {
   \lcA{\lcA{f}{\bFF}}{\lcA{\lcA{p}{f}}{x}}
  }
 }$.
\item[$\wSucO$] is
$\lcF{p}
 {
  \lcF{f x}
  {
   \lcA{\lcA{f}{\bTT}}{\lcA{\lcA{p}{f}}{x}}
  }
 }$.
\item[$\wXorBW$] is
$\lcF{l m}
     {
      \lcA{\wMap{\lcPD{a}{b}
                      {\lcA{\lcA{\bXor}{a}}{b}}}
          }
          {
           \lcA{\lcA{\wMT}{l}}{m}
          }
     }$.
\end{description}
\fi

\end{document}